\documentstyle[12pt,aasms]{article}

\newcommand{\bq}{\begin{equation}}
\newcommand{\eq}{\end{equation}}

\begin{document}

\title{A Precise Distance Indicator:\\
     Type Ia Supernova Multicolor Light Curve Shapes}

\author{Adam G. Riess, William H. Press, Robert P. Kirshner}
\affil{Harvard-Smithsonian Center for Astrophysics, 60 Garden Street, Cambridge,
 MA 02138}

\begin{abstract}
       We present an empirical method that uses multicolor light curve shapes (MLCS) to estimate the luminosity, distance, and total line-of-sight extinction of Type Ia supernovae (SN Ia).  The empirical correlation between the MLCS and the luminosity is derived from a ``training set'' of nine SN Ia light curves with independent distance and reddening estimates.  We find that intrinsically dim SN Ia are redder and have faster light curves than the bright ones which are slow and blue.  By thirty-five days after maximum the intrinsic color variations become negligable.   A formal treatment of extinction employing Bayes' theorem is used to estimate the best value and its uncertainty.    Applying MLCS to both light curves and to color curves provides enough information to determine which supernovae are dim because they are distant, which are intrinsically dim, and which are dim because of extinction by dust. The precision of the MLCS distances is examined by constructing a Hubble diagram with an independent set of twenty SN Ia's.  The dispersion of 0.12 mag indicates a typical distance accuracy of $5\%$ for a single object, and the intercept yields a Hubble constant on the Cepheid distance scale (Sandage et al 1994, 1996) of $H_0=65 \pm$ 3 (statistical) km s$^{-1}$ Mpc$^{-1}$ ($ \pm 6 $ total  error).  The slope of 0.2010 $\pm$ 0.0035 mag over the distance interval 32.2 $< \mu <$ 38.3 yields the most precise confirmation of the linearity of the Hubble law.  
\end{abstract}
subject headings:  supernovae:general ; cosmology: distance scale
\vfill
\eject
\section{Inhomogeneity of Type Ia Supernovae}

     Since the Curtis-Shapley debate of 1920 (Curtis 1921, Shapley 1921), the determination of supernova (SN) luminosities has been central to the discussion of extragalactic distances.  Shapley (1919) argued against the ``Island Universe'' hypothesis, because it required certain novae (such as S Andromedae of 1885) to reach the astonishing luminosity of $M=-16$.  He considered this to be ``out of the question''.   Curtis (1921) countered, concluding that ``the dispersion of the novae in spirals and in our galaxy may reach ten absolute magnitudes...a division into two classes is not impossible''.  This distinction between novae and supernovae, required by the extragalactic nature of the nebulae,  was later made explicit by Baade and Zwicky (1934).  They showed that in addition to their tremendous difference in absolute luminosity, the photometric and spectroscopic behavior of supernovae is distinct from novae.  Baade (1938) showed that supernovae were more uniform than novae, with a dispersion at peak of  1.1 magnitudes, making them suitable as extragalactic distance indicators.  

     The precision of supernova distance estimates has increased as the SN Ia class has been better understood and more narrowly defined.  The low dispersion in Baade's sample benefited from the fortuitous absence of Type II supernovae, which are significantly less luminous in the mean.  Beginning with SN 1940B, Type II supernovae were classified by the presence of hydrogen in their spectra (Minkowski 1941). The growing list of spectroscopically defined Type I supernovae had dispersions at peak of 0.8-0.6 magnitudes (Minkowski 1964, Kowal 1968, Kirshner et al. 1973, Oke and Searle 1974).  However, this sample included a number of ``peculiar'' SN Ia noted for their lack of silicon, which are now recognized to arise from massive stars that lose their envelope before core collapse (see Wheeler \& Harkness 1990).  After removing these silicon-deficient objects, now classified as Ib's and Ic's (Doggett and Branch 1985, Uomoto and Kirshner 1985, Wheeler and Levreault 1985, Wheeler and Harkness 1986), the remaining Type Ia supernovae (SN Ia)  form a more homogeneous set which serve as even more precise indicators of astronomical distances.  Leibundgut (1989) devised a set of standard templates to describe the photometric behavior of SN Ia and to estimate the peak apparent magnitude.  Hubble diagrams constructed using the peak of photographic SN Ia light curves had observed dispersions ranging from $\sigma_M$=0.65-0.36 magnitudes depending on which objects and color bands were used (Tammann and Leibundgut 1990, Branch and Miller 1993, Miller and Branch 1990, Della Valle and Panagia 1992, Rood 1994, Sandage and Tammann 1993, Sandage et al 1992,1993,1994). 
 An ambitious program to calibrate nearby SN Ia through Cepheid variables observed with the Hubble Space Telescope has been undertaken by Sandage et al (1992, 1993, 1994, 1996).  Assuming SN Ia to be homogeneous yields a Hubble constant in the range 50-58 km  s$^{-1}$  Mpc$^{-1}$ (Sandage et al 1992,1993,1994, 1995, Schaefer 1994, 1995a,b, 1996,  Branch \& Tammann 1992) with the most recent measurement giving 57 $\pm \ 4$ km  s$^{-1}$  Mpc$^{-1}$.  We show in \S 6 and \S 7 that the precision of this measurement is improved and the value of H$_0$ altered by including information contained in the light and color curve shapes.

 The hypothesis that SN Ia are standard candles drew support not only from empirical studies, but also from the earliest theoretical models which suggested they arose from ignition of a carbon-oxygen white dwarf at the Chandrasekhar mass (Hoyle and Fowler 1960, Arnett 1969, Colgate \& McKee 1969).  In these models a supersonic shock wave travels through the degenerate star, burning material into  $^{56}$Ni at a temperature of five billion degrees  (Khokhlov, Muller, and H\"{o}flich 1993, Mazurek \& Wheeler 1980).  Because the detonation is supersonic, the pre-shock region cannot expand to decrease the pressure or the burning temperature.  Further, the Fermi pressure of the degenerate material in the post-shock region remains insensitive to temperature for longer than the burning timescale.  The result is a total incineration and the production of a pure mass of nickel.  Such a standard explosion of a uniform mass would lead to a homogeneous light curve and uniform luminosity.  Yet, these complete burning models of Ia's do not reproduce the intermediate mass elements which are seen in the spectra of SN Ia (Wheeler and Harkness, 1990).  A successful model (Nomoto, Thielemann, and Yokoi 1984) which matched the observational constraints was so persuasive that Arnett, Branch, and Wheeler (1985) and Branch (1992) suggested calibration of the Hubble constant based only on theoretical models of uniform nickel production.  However, a variety of models  (Livne 1990, Khokhlov, Muller, and  H\"{o}flich 1993,  Woosley \& Weaver 1992, H\"{o}flich, Khokhlov, \& Wheeler 1995) match the observed features of the spectra and produce a range of nickel masses, and a range of predicted luminosities.   These models employ subsonic deflagration fronts, pulsations, or off-center explosions to allow the surface layers to pre-expand and burn at low temperature.  The success of these models in reproducing the observed spectra opens a large range of theoretical possibilities.  Unlike the first monoenergetic models, these models suggest a wide range of luminosities might result from the ignition of a white dwarf.

  Recently, precise observations of SN Ia made with CCD detectors show evidence for inhomogeneity in both luminosity and light curve shape.  One of the first SN Ia observed with a distinctly different light curve was 1986G (Phillips et al 1987, Cristiani et al 1992) which displayed a spectacularly rapid decline in its B and V light curve  and unique spectral characteristics including stronger-than-usual Si features.  Although SN 1986G was heavily reddened by dust, reddening cannot significantly alter the {\it shape} of the light curve.  SN 1991bg in NGC 4374 is the most extreme SN Ia in an increasingly apparent photometric and spectral sequence.  Leibundgut et al (1993) (also Filippenko et al 1992) described a number of photometric abnormalities of 1991bg with respect to his templates.  These include the fastest post-maximum decline (2.05 and 1.42 mag decrease drop in B and V in the fifteeen days after maximum compared to 1.22 and 0.64 mag for the templates in B and V), a narrow luminosity peak, and an intrinsic red color near maximum. A simple and convincing argument that SN Ia have a large spread in luminosity is that SN 1957B, which occurred in the same galaxy, was 2.5 magnitudes brighter in B than SN 1991bg.  In addition, SN 1991bg was at least 2 magnitudes fainter than other SN Ia in the Virgo cluster, of which NGC 4374 is a member.  This extreme SN Ia seems to have a twin in the sub-luminous SN 1992K (Hamuy et al 1994), which strongly resembles 1991bg in photometric and spectral behavior.  

At the opposite extreme of the SN Ia class, SN 1991T showed spectral and photometric peculiarities which were different from those seen in the rapid decliners.  Phillips et al (1992) found the light curves in B and V to rise and decline more slowly than the standard templates near maximum, and a month after peak, this shallower decline resulted in a light curve 0.2-0.3 mag brighter than the templates.  Although SN 1991T's host galaxy, NGC 4527, appears to lie south of the main Virgo cluster,  Phillips et al (1992, 1993) estimates the peak luminosity exceeded that of other SN Ia in Virgo by 0.3-0.5 mag.  From the narrow Na I D absorption line, Filippenko et al (1992) deduced that SN 1991T is dimmed by dust in NGC 4527 and concluded it may have been as much as
 $ \sim 0.9 $ mag more luminous than a typical SN Ia.  More recently, the Cal\'{a}n/CTIO supernova search yielded SN 1992bc and 1992bo (Maza et al 1994), two SN Ia with similar recession velocities of 6050 and 5660 km s$^{-1}$, but with peak apparent luminosities differing by
0.69 mag in B.  The large difference in apparent magnitude and the small difference in recession velocity imply that SN 1992bc was intrinsically brighter by $0.8\pm0.2$ mag than 1992bo.  For a difference in distance to account for this difference in luminosity, the peculiar velocities of the two supernovae would have to differ by $\sim 2,500$ km sec$^{-1}$, an unlikely alternative.  Interestingly, SN 1992bc declined more slowly than the average template while SN 1992bo's post-maximum fall was more rapid than the average template, a result in good accord with later analysis by Phillips (1993).
  
Even before these recent examples of inhomogeneity, less precise photographic measurements by Barbon et al (1973) suggested that there might exist two photometric classes; those with ``fast'' decline rates after maximum which were intrinsically brighter supernovae, and those with ``slow'' decline rates which were fainter.   With the poor quality of photographic and photoelectric photometry then available, such a real distinction was difficult to demonstrate convincingly.   Pskovskii (1977,1984) suggested a continuous photometric sequence of SN Ia light curves.  He introduced a parameter, $\beta$: the slope of the B band post-maximum decline in magnitudes per 100 days.  Using 54 photographic SN Ia light curves, Psovskii found a weak correlation between $\beta$ and the absolute B magnitude at maximum light which was {\it opposite} to that of Barbon (1973).   Early difficulties in measuring a relation between SN Ia light curve shape and intrinsic luminosity resulted from noisy photographic data which were poorly sampled and calibrated.  These difficulties were compounded by the problem of measuring a decaying light curve on a bright galaxy with a non-linear photographic detector (Boisseau and Wheeler 1991).  With the advantage of better data measured with linear detectors, Phillips (1993) demonstrated conclusive evidence for a luminosity-light curve decline relation among SN Ia.  Using a set of well-sampled SN Ia light curves with precise optical photometry and accurate relative distances, Phillips found that the absolute luminosity in B,V, and I is correlated with the B band decline in the fifteen days following maximum light.  The sense of the correlation is that dimmer SN Ia fall more rapidly after maximum than bright SN Ia.  Application of this relation to his sample results in a significant reduction of the dispersion in B,V, and I luminosity from 0.8, 0.6, and 0.5 mag to 0.36, 0.28, and 0.38 respectively.  A more recent investigation by Tammann and Sandage (1995)  has examined the luminosity-light curve decline relation among ``normal'' SN Ia, in this case ``normal'' is defined as having a $(B-V)_{max}$ $\leq 0.30$ mag.  They tentatively confirm a luminosity dependence on light curve decline, with a slope that is shallower than Phillips (1993) but consistent with Hamuy et al (1995) for a similar set of ``normal'' SN Ia.  In \S 6 we show for a sample restricted to ``normal'' SN Ia, a significant correlation between light curve shape and luminosity exists.

 This empirical relation between light curves and luminosity in SN Ia has been paced by an abundance of theoretical models which can account for the observed behavior (H\"{o}flich, Khokhlov, and Muller 1993, Woosley and Weaver 1994, Ruiz-Lapuente et al 1993, Livne and Arnett 1995, H\"{o}flich and Khokhlov 1995).  These new models include deflagration burning fronts, off-center detonations, surface helium burning, pulsed delayed detonations and sub-Chandrasekhar progenitors.  These models give plausible causes for the observed inhomogeneity of SN Ia and for the origin of the empirical relations between light curve shape and luminosity.

 Most recently, Hamuy et al (1995) have employed a template-fitting approach and we (Riess, Press, and Kirshner 1995a, hereafter RPK 95a) have developed a linear estimation algorithm to use the distance-independent light curve shapes to improve the precision of distance measurements to SN Ia.  The techniques have much in common and both yield Hubble diagram dispersions of $\sim$ 0.2 magnitudes for an overlapping set of objects.  The Light Curve Shape (LCS) method, described in RPK 95a, has the advantage of providing quantitative error estimates for the distance.  The present paper extends the LCS technique to use the shapes of B$-$V, V$-$R, and V$-$I color curves which provide enough information to determine the relation between absolute luminosity and intrinsic color.  Knowledge of an SN Ia's intrinsic color allows us to measure the extinction from the observed reddening.   For each well-observed SN Ia we measure the luminosity, extinction and the extinction-corrected distance.  The multicolor light curve shape (MLCS) method significantly increases the precision of distance estimates from SN Ia light curves as we show in \S 6.

There are many potential applications for a bright distance indicator with $<$ $10 \%$ precision.  Nearby $(0.01 \leq z \leq 0.1)$, it should be possible to measure the Hubble constant to an accuracy which is limited only by the underlying calibration of Cepheids.    It is important to compare the Hubble constant derived using the light curve shape-luminosity relation with determinations which have assumed a homogeneous luminosity for SN Ia (Sandage et al. 1992, 1993, 1994, 1996).  Using the velocity residuals from Hubble flow, we have measured the motion of the Local Group with respect to the rest frame of galaxies with supernovae (Riess, Press, and Kirshner 1995b).  At even greater distances $(0.3 \leq z \leq 0.6)$ MLCS distance measurements of all well-observed SN Ia could be used to determine the cosmological deceleration parameter, $q_o$ (Perlmutter et al 1995, Perlmutter et al 1996, Schmidt et al 1996, IAUC 6160).

  Some have sought to improve the homogeneity of the observed SN Ia by restricting the sample to supernovae with ``normal'' spectra (Branch, Fisher, and Nugent 1993, Sandage et al 1994, 1996).  While spectroscopic information may prove useful in producing smaller dispersion in distance estimates,  the difficulty in obtaining good spectra of very distant SN Ia makes it hard to identify subtle spectral variations. A sample cut which cannot be applied with equal effectiveness for nearby and for distant SN Ia could lead to a bias in cosmological parameters determined by them.  We prefer to develop a method that can be applied to all SN Ia.

Given at least one light curve and one color curve, photoelectrically observed within ten days after maximum, our MLCS method can distinguish between the effects of distance, intrinsic dimness, and dust for all SN Ia.  In \S 2-5 we develop the multicolor light curve shape method for measuring extinction-corrected distances.  In \S 6, for an independent sample, we compare the extinction-corrected MLCS distances with distances determined by more limited assumptions.  In \S 7, we estimate the Hubble constant and discuss sample membership, selection bias and the range of progenitors which may be responsible for the empirical range of SN Ia luminosities and colors

\section{Learning Curves}

SN Ia light curves are noisy unevenly sampled time series, which, when diligently observed, are available in four bands, B, V, R, and I.  Although some SN Ia are found in ellipticals, most are found in spirals and irregulars where they can be subject to significant extinction by dust.  At present, the number of well sampled SN Ia light curves available on a modern photometric system is limited (i.e. $\leq$ 50) (Hamuy et al 1995, Riess et al 1996, Ford et al 1993).  

  We employ a general $\chi^2$ model to establish the empirical relation between light and color curve shape and luminosity from a nearby subset of SN Ia with accurately known {\it relative distances} and extinctions.  Then we use these results to estimate the extinction-corrected distance for a separate distant sample {\it solely from the observed light and color curves}.  The observables are a visual band light curve and up to three color curves.  

To first approximation, the observed light and color curves of SN Ia are homogeneous, resembling standard template light and color curves with the addition of noise ({\bf n}) and individual offsets that result from the apparent distance modulus ($\mu$) and a color excess ($E_{color}$) due to reddening by dust:

\bq {\bf m_V} = {\bf M_V} + \mu + {\bf n} \eq

\bq {\bf m_B-m_V} = {\bf (B-V)}_0 + E_{B-V} + {\bf n} \eq      

 \bq {\bf m_V-m_R} = {\bf (V-R)}_0 + E_{V-R} + {\bf n} \eq      

\bq {\bf m_V-m_I} = {\bf (V-I)}_0 + E_{V-I} + {\bf n} \eq      

Here bold-faced denotes a vector whose entries are measured or determined as a function of time.  In standard convention, $M$ is an absolute magnitude, $m$ is an apparent magnitude, and $(B-V)_0$, $(V-R)_0$, and $(V-I)_0$ are unreddened color curves.   Each available color curve yields an $E_{color}$ term which, combined with a standard extinction curve, reduces the uncertainty in measuring the total line-of-sight extinction, $A_V$.  This description of light and color curves assumes intrinsic homogeneity of SN Ia, and is equivalent to the ``template'' approach of Leibundgut (1989), Sandage et al (1994, 1996), and Sandage and Tammann (1993).

 Improvements in the quality of the available data set and the success of Phillips (1993) motivate an approach which, to first order, can account for the observed variations in the light and color curves and intrinsic luminosity.  The most economical approach is to adopt a single parameter and correlate it with the variations of the observed curves. The natural parameter to choose is the amount by which the intrinsic luminosity differs from an SN Ia of standard brightness and curve shape since, in the end, it is this difference we hope to measure.  We call this the ``luminosity correction'', $\Delta$, where $\Delta \equiv M_V-M_{V,standard}$, and $M_{V,standard}$ is the luminosity of the SN Ia chosen to depict the ``standard'' SN Ia event.  This difference is measured by convention at the date of B maximum.  
The functions of time which,  to first order in $\Delta$, correct the observed curves to the template curves are ``correction templates'', $R_V(t)$ or $R_{color}(t)$.  Using these definitions, we expect two supernovae whose absolute visual luminosities differ by $\Delta$ magnitudes to have light and color curves which differ by $R_V(t) \Delta$ and $R_{color}(t) \Delta$.  The assumption that luminosity correlates {\it linearly} with light and color curve shape need not provide a complete description, but given the limitations of our real data sets, it is a reasonable way to begin.  The evidence that this is a useful way to proceed is provided in \S 6 where we show that this method produces a significant decrease in the dispersion for an independent set of SN Ia in the Hubble flow.   

     Our improved model for the light and color curves is thus
\bq {\bf m_V} = {\bf M_V} + \mu +  {\bf R_V} \Delta +{\bf n} \eq where $R_V(t) \Delta$ gives the deviation from the template, $M_V(t)$, at each time.  Similarly,

\bq {\bf m_B-m_V} = {\bf (B-V)}_0 + E_{B-V} + {\bf R_{B-V}} \Delta + {\bf n} \eq

\bq {\bf m_V-m_R} = {\bf (V-R)}_0 + E_{V-R} + {\bf R_{V-R}} \Delta + {\bf n} \eq

\bq {\bf m_V-m_I} = {\bf (V-I)}_0 + E_{V-I} + {\bf R_{V-I}} \Delta + {\bf n} \eq
 
where $(B-V)_0$, $(V-R)_0$, and $(V-I)_0$ are the unreddened template color curves.  It is instructive to compare equations (5) through (8) which allow SN Ia inhomogeneity to equations (1) through (4) which impose SN Ia homogeneity.

  In matrix format we can write equations (5) through (8) as one system of equations   

\bq \pmatrix{m_V(t_1) \cr m_V(t_2) \cr.\cr m_V(t_N) \cr \cr m_B(t_1)-m_V(t_1
) \cr m_B(t_2)-m_V(t_2) \cr .  \cr m_B(t_N)-m_V(t_N) \cr \cr m_V(t_1)-m_R(t_1) \cr m
_V(t_2)-m_R(t_2) \cr .  \cr m_V(t_N)-m_R(t_N) \cr \cr m_V(t_1)-m_I(t_1) \cr m
_V(t_2)-m_I(t_2) \cr .  \cr m_V(t_N)-m_I(t_N)} =
\pmatrix{M_V(t_1) \cr M_V(t_2)\cr.\cr M_V(t_N) \cr \cr (B-V)_0(t_1)
 \cr (B-V)_0(t_2) \cr . \cr (B-V)_0(t_N) \cr \cr (V-R)_0(t_1) \cr (V-R)_0(t_2) \cr .
 \cr (V-R)_0(t_N) \cr \cr (V-I)_0(t_1) \cr (V-I)_0(t_2) \cr .
 \cr (V-I)_0(t_N)} \ \ \ \ \ + \ \ \ \pmatrix{1
&0&R_V(t_1)\cr1&0&R_V(t_2)\cr.&.&.\cr1&0&R_V(t_N) \cr \cr 0&{1
\over 3.1}&R_{B-V}(t_1)
\cr0&{1 \over 3.1} &R_{B-V}(t_2)\cr.&.&.\cr0&{1 \over 3.1}&R_{B-V}(t_N) \cr \cr 0&{1
\over 3.9}&R_{V-R}(t_1)\cr0&{1 \over 3.9}
&R_{V-R}(t_2)\cr.&.&.\cr0&{1 \over 3.9}&R_{V-R}(t_N) \cr \cr 0&{1 \over 1.9}&R_{V-I}(
t_1) \cr 0&{1 \over 1.9}&R_{V-I}(t_2) \cr .&.&. \cr 0&{1 \over 1.9}&R_{V-I}(t_N)
 }\ \ \ \ * \ \ \ \ \pmatrix{\mu \cr A_V \cr \Delta} \ + n(t)\eq where we have combined the $E_{color}$ terms into one simultaneous measurement of $A_V$ with the aid of the standard reddening law parameterization, ${ A_V \over E(B-V)}=3.1, { A_V \over E(V-R)}=3.9,{ A_V \over E(V-I)}=1.9$ (Savage \& Mathis, 1979).  Here, we assume the Galactic reddening ratios are valid for the dust in distant galaxies.  Elsewhere, we examine the homogeneity of  these ratios in distant galaxies, and find the most likely form of the reddening law for dust in distant galaxies is consistent with the Galactic reddening law (Riess, Press, Kirshner 1996b). 
      
   The $\chi^2$ of the fit between data and model is \bq \chi^2=({\bf n})^T {\bf C^{-1}} ({\bf n}) \eq where {\bf n} is the vector of residuals in equations (5) through (8) or equivalently in equation (9).   Here,  {\bf C} is the correlation matrix whose elements are intended to reflect the errors in observations and residual, unmodeled correlations in SN Ia light and color curves.   The correlation matrix is discussed in detail in \S 4.
    
 Rybicki and Press (1992) have derived two analytical minimizations of the $\chi^2$ given in equation (10).  One gives the best estimate of the correction templates, $R_{V}(t)$ or $R_{color}(t)$, provided the SN Ia parameters, $[\mu \ A_V \ \Delta]$, are known, the other gives the best estimate of the parameters provided the correction templates are known.  We employ both, using the nearby set with accurate relative distances as a ``training set'' to estimate $R_V(t)$ and $R_{color}(t)$.  Once trained, we measure the $\mu$, $A_V$, and $\Delta$ parameters for an independent set of SN Ia from the shapes of their light and color curves.

  \begin{table}[tb]   
\begin{center}
\caption{Training Set} \vspace{0.4cm}
\begin{tabular}{ccccccccc} \hline \hline
SN & Galaxy & $ \mu(\sigma) $ & Method & $ \mu  Ref $ & $E(B-V)(\sigma)$ & $ M_V $ & $\Delta(\sigma)$ & $Phot.
 Ref$ \\      
\hline
1980N & N1316 & 31.15(0.10) & SBF,PNLF & 2 & 0.00(0.02) & -18.71 & -0.17(0.10) & 1 \\
1981B & N4536 & 30.50(0.30) & T-F & 1 & 0.00(0.02) & -18.54 & 0.00(0.32) & 2 \\
1986G & N5128 & 27.72(0.10) & SBF,PNLF & 2 & 0.60(0.10) & -18.13 & 0.41(0.31) & 3 \\
1989B & N3627 & 29.40(0.30) & T-F & 1 & 0.35(0.03) & -18.50 &0.04(0.31)& 4 \\
1990N & N4639 & 31.40(0.30) & T-F & 1 & 0.01(0.02) & -18.82 &-0.28(0.31)& 5 \\
1991T & N4527 & 30.60(0.30) & T-F & 1 & 0.00(0.02) & -19.10 &-0.56(0.31)& 6,7 \\
1991bg & N4374 & 31.05(0.10) & SBF,PNLF & 2 & 0.00(0.02) & -17.10 &1.44(0.10)& 8,9 \\
1992A & N1380 & 30.65(0.11) & SBF & 4 & 0.00(0.02) & -18.10 &0.44(0.13)& 10 \\
1994ae & N3370 & 31.28(0.30) & T-F & 5 & 0.14(0.03) & -18.65 &-0.13(0.30)& 11 \\
\hline \hline 
\multicolumn{8}{l}{ \scriptsize \ \ $\mu$ references-(1) Pierce 1994; (2) Ciardullo, Jacoby, \& Tonry 1993;} \\
\multicolumn{8}{l}{ \scriptsize (4) Tonry 1991 (5) Dell Antonio 1995}\\
\multicolumn{8}{l}{ \scriptsize \ \ photometry references- (1) Hamuy {\it et al} 1991;(2) Buta and Turner 1983; (3) Phillips {\it et al} 1987}\\
\multicolumn{8}{l}{ \scriptsize (4) Wells {\it et al} 1993;(5) Leibundgut {\it et al} 1991; (6) Phillips {\it et al} 1992;(7) Ford {\it et al} 1993;} \\
\multicolumn{8}{l}{\scriptsize(8) Filippenko {\it et al} 1992;(9) Leibundgut {\it et al} 1993; (10) Suntzeff {\it et al} 1996 (11) Riess et al 1996} \\

\normalsize

\end{tabular}
\end{center}
\end{table}
     Analytical minimization of $\chi^2$ in equation (10) with respect to $R_V(t)$ or $R_{color}(t)$ is independent of {\bf C}
and gives;
  
\bq R_V(t)={\langle [m_V(t)-M_V(t)-\mu]\Delta \rangle \over \langle {\Delta}^2 \rangle } \eq
or
\bq R_{color}(t)={\langle [m_{color}(t)-M_{color}(t)-E_{color}]\Delta \rangle \over \langle {\Delta}^2 \rangle } \eq  where the angle brackets denote an average over the training set weighted by the uncertainty in the given estimates of $\Delta$. 

 It is convenient to choose a particular object as the ``standard'' SN Ia whose light and color curves define the standard template curves, $M_V(t)$ and $M_{color}(t)$.   By definition, the  ``standard'' SN Ia have a peak luminosity variation of $\Delta \equiv$0.  The utility of the method is insensitive to the choice of the template curves and luminosity since all quantities are defined relative to it.    Leibundgut's templates, made from SN 1989B, SN 1980N, and SN 1981B approximate the light curves of the ``normal'' training set supernovae in B and V.  For R and I, we have constructed our own templates from SN 1989B, SN 1980N, and SN 1981B. 

In solving for the correction templates, we only need consistent {\it relative} distances for our training set of SN Ia since we correlate luminosity {\it differences} with light curve shape variation.   No choice of absolute distance scale is necessary since all luminosity corrections, $\Delta$, are relative to the standard SN Ia.  Surface Brightness Fluctuations (SBF), Planetary Nebula Luminosity Function (PNLF), and Tully and Fisher's luminosity-line width relation (T-F) provide accurate and consistent bias-corrected {\it relative} distances to the host galaxies of the training set of SN Ia (Strauss \& Willick 1995,  Jacoby et al 1992, Ciardullo, Jacoby, and Tonry 1993, Pierce 1994, Kennicutt, Freedman, and Mould 1995).  We have no dogma about the correctness of this {\it absolute} distance scale--in \S 7 we calibrate the absolute luminosity of a standard SN Ia, contained in $M_V(t)$ with the Cepheid observations from Sandage et al (1994, 1996).  This approach has the advantage of placing our distance scale directly onto that of the Cepheid variables.

The training set of supernovae we have used also has precise optical photometry, well-sampled light curves, and estimates of $E(B-V)$ (see table 1).  The color excesses  in table 1 are estimated from comparisons with unreddened, photometrically similar SN Ia (Phillips 1993, Wells et al 1994) and are used to correct the luminosity and color curves of the affected objects.  In \S 7 we discuss our training set membership.  On the SBF-PNLF-TF distance scale, SN Ia which define the template have an $<M_V>$=--18.54 on the date of B maximum; therefore each supernova's $\Delta \equiv M_V-(-18.54)$.  We have calculated the correction templates from equations (11) and (12) using the training set of supernovae provided in table 1.  The correction templates and standard templates are listed in table 2. The standard template, $M_V(t)$, as listed in table 2, is calibrated on the Cepheid distance scale discussed in \S 7.

\begin{table} 
\caption{Training Vectors}
\begin{tiny}  
\begin{center}
\begin{tabular}{ccccccccccccc}
\hline
\hline
\multicolumn{1}{c}{\small phase} & \multicolumn{4}{c}{\small standard templates} & \multicolumn{4}{c}{\small correction templates} & \multicolumn{4}{c}{\small weighting vectors} \\
\hline  
$ t_{Bmax}$ &  $M_V(t)$ &  $M_{B-V}(t)$ &  $M_{V-R}(t)$ &  $M_{V-I}(t)$ &  $R_V(t)$ &  $R_{B-V}(t)$ &  $R_{V-R}(t)$ &  $R_{V-I}(t)$ &  $\sigma_V(t)$ &  $\sigma_B(t)$ &  $\sigma_R(t)$ &  $\sigma_I(t)$ \\
\hline
  -11.000   &  -18.061   &   -0.307   &    0.007   &    0.390   &    1.257   &    0.492   &    0.273   &    0.408   &    0.000   &    0.371   &    0.130   &    0.251  \\
  -10.000   &  -18.374   &   -0.326   &    0.009   &    0.294   &    1.257   &    0.493   &    0.273   &    0.408   &    0.002   &    0.366   &    0.112   &    0.221  \\
   -9.000   &  -18.534   &   -0.285   &    0.012   &    0.260   &    1.257   &    0.493   &    0.273   &    0.408   &    0.008   &    0.349   &    0.112   &    0.216  \\
   -8.000   &  -18.693   &   -0.244   &    0.015   &    0.227   &    1.259   &    0.494   &    0.273   &    0.408   &    0.011   &    0.332   &    0.111   &    0.211  \\
   -7.000   &  -18.824   &   -0.216   &    0.019   &    0.185   &    1.259   &    0.491   &    0.272   &    0.404   &    0.013   &    0.316   &    0.114   &    0.212  \\
   -6.000   &  -18.955   &   -0.189   &    0.022   &    0.144   &    1.253   &    0.489   &    0.271   &    0.401   &    0.012   &    0.299   &    0.116   &    0.214  \\
   -5.000   &  -19.105   &   -0.102   &    0.029   &    0.002   &    1.249   &    0.479   &    0.268   &    0.407   &    0.011   &    0.284   &    0.119   &    0.219  \\
   -4.000   &  -19.183   &   -0.081   &    0.035   &   -0.079   &    1.244   &    0.474   &    0.257   &    0.408   &    0.010   &    0.270   &    0.123   &    0.226  \\
   -3.000   &  -19.246   &   -0.060   &    0.042   &   -0.157   &    1.220   &    0.473   &    0.257   &    0.412   &    0.008   &    0.258   &    0.125   &    0.233  \\
   -2.000   &  -19.296   &   -0.040   &    0.049   &   -0.216   &    1.160   &    0.471   &    0.257   &    0.412   &    0.007   &    0.248   &    0.128   &    0.241  \\
   -1.000   &  -19.334   &   -0.020   &    0.051   &   -0.267   &    1.093   &    0.468   &    0.258   &    0.412   &    0.006   &    0.239   &    0.130   &    0.249  \\
    0.000   &  -19.360   &   -0.000   &    0.046   &   -0.312   &    1.029   &    0.470   &    0.256   &    0.417   &    0.005   &    0.233   &    0.131   &    0.256  \\
    1.000   &  -19.375   &    0.021   &    0.034   &   -0.353   &    1.002   &    0.531   &    0.256   &    0.525   &    0.006   &    0.229   &    0.131   &    0.262  \\
    2.000   &  -19.380   &    0.043   &    0.014   &   -0.388   &    1.014   &    0.550   &    0.260   &    0.499   &    0.007   &    0.226   &    0.131   &    0.268  \\
    3.000   &  -19.375   &    0.068   &   -0.013   &   -0.418   &    1.033   &    0.576   &    0.256   &    0.615   &    0.008   &    0.226   &    0.130   &    0.273  \\
    4.000   &  -19.362   &    0.095   &   -0.042   &   -0.441   &    1.035   &    0.616   &    0.262   &    0.612   &    0.011   &    0.226   &    0.128   &    0.276  \\
    5.000   &  -19.340   &    0.125   &   -0.069   &   -0.458   &    1.060   &    0.668   &    0.291   &    0.658   &    0.014   &    0.228   &    0.126   &    0.279  \\
    6.000   &  -19.311   &    0.158   &   -0.091   &   -0.469   &    1.102   &    0.683   &    0.323   &    0.704   &    0.018   &    0.231   &    0.124   &    0.280  \\
    7.000   &  -19.275   &    0.194   &   -0.106   &   -0.473   &    1.142   &    0.734   &    0.372   &    0.747   &    0.022   &    0.235   &    0.121   &    0.280  \\
    8.000   &  -19.233   &    0.234   &   -0.115   &   -0.473   &    1.186   &    0.705   &    0.396   &    0.800   &    0.027   &    0.240   &    0.119   &    0.279  \\
    9.000   &  -19.187   &    0.277   &   -0.115   &   -0.468   &    1.233   &    0.675   &    0.418   &    0.851   &    0.033   &    0.245   &    0.116   &    0.277  \\
   10.000   &  -19.136   &    0.323   &   -0.109   &   -0.458   &    1.274   &    0.651   &    0.411   &    0.861   &    0.039   &    0.251   &    0.114   &    0.274  \\
   11.000   &  -19.082   &    0.372   &   -0.095   &   -0.443   &    1.317   &    0.629   &    0.406   &    0.870   &    0.045   &    0.256   &    0.111   &    0.270  \\
   12.000   &  -19.025   &    0.423   &   -0.075   &   -0.422   &    1.358   &    0.607   &    0.402   &    0.881   &    0.052   &    0.262   &    0.109   &    0.266  \\
 13.000   &  -18.967   &    0.476   &   -0.051   &   -0.394   &    1.396   &    0.583   &    0.397   &    0.893   &    0.058   &    0.267   &    0.107   &    0.261  \\
   14.000   &  -18.906   &    0.530   &   -0.022   &   -0.361   &    1.468   &    0.629   &    0.388   &    0.904   &    0.065   &    0.272   &    0.106   &    0.255  \\
   15.000   &  -18.845   &    0.584   &    0.010   &   -0.321   &    1.482   &    0.578   &    0.382   &    0.915   &    0.072   &    0.276   &    0.105   &    0.249  \\
   16.000   &  -18.783   &    0.638   &    0.044   &   -0.268   &    1.497   &    0.523   &    0.378   &    0.926   &    0.079   &    0.280   &    0.105   &    0.243  \\
   17.000   &  -18.721   &    0.692   &    0.078   &   -0.204   &    1.507   &    0.475   &    0.373   &    0.936   &    0.085   &    0.282   &    0.105   &    0.237  \\
   18.000   &  -18.660   &    0.745   &    0.112   &   -0.130   &    1.500   &    0.437   &    0.362   &    0.907   &    0.092   &    0.284   &    0.105   &    0.231  \\
   19.000   &  -18.598   &    0.796   &    0.146   &   -0.047   &    1.493   &    0.397   &    0.329   &    0.792   &    0.098   &    0.285   &    0.106   &    0.226  \\
   20.000   &  -18.537   &    0.844   &    0.177   &    0.043   &    1.487   &    0.352   &    0.294   &    0.690   &    0.104   &    0.285   &    0.107   &    0.220  \\
   21.000   &  -18.476   &    0.888   &    0.207   &    0.133   &    1.486   &    0.307   &    0.259   &    0.646   &    0.109   &    0.284   &    0.109   &    0.215  \\
   22.000   &  -18.415   &    0.930   &    0.233   &    0.220   &    1.493   &    0.270   &    0.226   &    0.601   &    0.115   &    0.282   &    0.111   &    0.211  \\
   23.000   &  -18.355   &    0.968   &    0.256   &    0.297   &    1.506   &    0.255   &    0.190   &    0.551   &    0.119   &    0.279   &    0.114   &    0.207  \\
   24.000   &  -18.295   &    1.002   &    0.276   &    0.365   &    1.488   &    0.207   &    0.159   &    0.502   &    0.124   &    0.275   &    0.116   &    0.204  \\
   25.000   &  -18.235   &    1.031   &    0.293   &    0.422   &    1.502   &    0.186   &    0.150   &    0.464   &    0.128   &    0.270   &    0.119   &    0.202  \\
   26.000   &  -18.176   &    1.056   &    0.306   &    0.469   &    1.479   &    0.166   &    0.140   &    0.426   &    0.131   &    0.264   &    0.123   &    0.200  \\
   27.000   &  -18.117   &    1.077   &    0.315   &    0.505   &    1.456   &    0.144   &    0.131   &    0.388   &    0.134   &    0.257   &    0.126   &    0.199  \\
   28.000   &  -18.059   &    1.094   &    0.321   &    0.537   &    1.439   &    0.122   &    0.122   &    0.350   &    0.137   &    0.250   &    0.129   &    0.199  \\
   29.000   &  -18.001   &    1.106   &    0.323   &    0.566   &    1.441   &    0.095   &    0.112   &    0.312   &    0.139   &    0.241   &    0.132   &    0.199  \\
   30.000   &  -17.944   &    1.114   &    0.323   &    0.591   &    1.439   &    0.076   &    0.103   &    0.273   &    0.141   &    0.233   &    0.136   &    0.200  \\
   31.000   &  -17.887   &    1.118   &    0.319   &    0.613   &    1.438   &    0.058   &    0.093   &    0.222   &    0.142   &    0.223   &    0.139   &    0.202  \\
   32.000   &  -17.832   &    1.119   &    0.315   &    0.631   &    1.436   &    0.040   &    0.084   &    0.158   &    0.144   &    0.214   &    0.142   &    0.205  \\
   33.000   &  -17.779   &    1.118   &    0.311   &    0.646   &    1.434   &    0.022   &    0.075   &    0.129   &    0.145   &    0.204   &    0.144   &    0.208  \\
   34.000   &  -17.726   &    1.113   &    0.307   &    0.657   &    1.431   &    0.005   &    0.066   &    0.109   &    0.145   &    0.193   &    0.147   &    0.211  \\
   35.000   &  -17.676   &    1.107   &    0.303   &    0.666   &    1.426   &   -0.011   &    0.061   &    0.094   &    0.145   &    0.183   &    0.149   &    0.215  \\
   36.000   &  -17.629   &    1.100   &    0.299   &    0.671   &    1.422   &   -0.026   &    0.064   &    0.090   &    0.146   &    0.173   &    0.150   &    0.219  \\
 37.000   &  -17.584   &    1.092   &    0.295   &    0.673   &    1.418   &   -0.041   &    0.067   &    0.086   &    0.145   &    0.163   &    0.152   &    0.223  \\
   38.000   &  -17.541   &    1.083   &    0.291   &    0.673   &    1.415   &   -0.052   &    0.070   &    0.081   &    0.145   &    0.153   &    0.153   &    0.228  \\
   39.000   &  -17.502   &    1.076   &    0.287   &    0.670   &    1.413   &   -0.049   &    0.073   &    0.076   &    0.145   &    0.144   &    0.153   &    0.232  \\
   40.000   &  -17.466   &    1.069   &    0.283   &    0.664   &    1.410   &   -0.047   &    0.075   &    0.072   &    0.145   &    0.135   &    0.154   &    0.237  \\
   41.000   &  -17.432   &    1.062   &    0.278   &    0.656   &    1.409   &   -0.045   &    0.077   &    0.069   &    0.144   &    0.127   &    0.153   &    0.241  \\
   42.000   &  -17.401   &    1.057   &    0.274   &    0.646   &    1.407   &   -0.042   &    0.078   &    0.067   &    0.144   &    0.119   &    0.153   &    0.245  \\
   43.000   &  -17.380   &    1.061   &    0.270   &    0.633   &    1.406   &   -0.040   &    0.080   &    0.065   &    0.143   &    0.112   &    0.152   &    0.249  \\
   44.000   &  -17.350   &    1.055   &    0.266   &    0.617   &    1.405   &   -0.038   &    0.081   &    0.063   &    0.143   &    0.106   &    0.151   &    0.252  \\
   45.000   &  -17.330   &    1.058   &    0.262   &    0.600   &    1.404   &   -0.035   &    0.083   &    0.060   &    0.143   &    0.100   &    0.149   &    0.255  \\
   46.000   &  -17.300   &    1.030   &    0.258   &    0.581   &    1.403   &   -0.033   &    0.084   &    0.058   &    0.142   &    0.095   &    0.148   &    0.257  \\
   47.000   &  -17.271   &    1.026   &    0.254   &    0.559   &    1.402   &   -0.031   &    0.086   &    0.062   &    0.142   &    0.091   &    0.146   &    0.258  \\
   48.000   &  -17.245   &    1.017   &    0.250   &    0.536   &    1.402   &   -0.031   &    0.091   &    0.134   &    0.142   &    0.088   &    0.143   &    0.259  \\
   49.000   &  -17.219   &    1.007   &    0.246   &    0.510   &    1.402   &   -0.031   &    0.097   &    0.165   &    0.143   &    0.086   &    0.141   &    0.259  \\
   50.000   &  -17.194   &    0.999   &    0.242   &    0.482   &    1.403   &   -0.028   &    0.099   &    0.198   &    0.143   &    0.084   &    0.139   &    0.258  \\
   51.000   &  -17.168   &    0.990   &    0.238   &    0.451   &    1.405   &   -0.035   &    0.103   &    0.224   &    0.143   &    0.084   &    0.136   &    0.257  \\
   52.000   &  -17.143   &    0.981   &    0.234   &    0.418   &    1.405   &   -0.028   &    0.107   &    0.138   &    0.144   &    0.084   &    0.134   &    0.254  \\
   53.000   &  -17.117   &    0.972   &    0.230   &    0.384   &    1.435   &   -0.042   &    0.108   &    0.194   &    0.145   &    0.085   &    0.132   &    0.251  \\
   54.000   &  -17.091   &    0.963   &    0.226   &    0.348   &    1.509   &   -0.069   &    0.109   &    0.252   &    0.146   &    0.087   &    0.130   &    0.247  \\
   55.000   &  -17.066   &    0.955   &    0.222   &    0.311   &    1.485   &   -0.063   &    0.108   &    0.309   &    0.147   &    0.090   &    0.128   &    0.243  \\
   56.000   &  -17.040   &    0.945   &    0.218   &    0.273   &    1.493   &   -0.060   &    0.105   &    0.351   &    0.148   &    0.093   &    0.126   &    0.237  \\
   57.000   &  -17.015   &    0.937   &    0.213   &    0.236   &    1.495   &   -0.061   &    0.083   &    0.351   &    0.150   &    0.097   &    0.125   &    0.231  \\
   58.000   &  -16.989   &    0.928   &    0.209   &    0.199   &    1.488   &   -0.065   &    0.058   &    0.344   &    0.151   &    0.101   &    0.124   &    0.224  \\
   59.000   &  -16.963   &    0.918   &    0.205   &    0.162   &    1.487   &   -0.079   &    0.039   &    0.356   &    0.153   &    0.105   &    0.124   &    0.217  \\
\hline    
\end{tabular}  
\end{center}  
\end{tiny}   
\end{table}

\begin{table}
\begin{tiny}
\begin{center}  
\begin{tabular}{c|cccc|cccc|cccc} 
\hline
t & $M_V(t)$ & $M_{B-V}(t)$ & $M_{V-R}(t)$ & $M_{V-I}(t)$ & $R_V(t)$ & $R_{B-V}(t)$ & $R_{V-R}(t)$ & $R_{V-I}(t)$ & $\sigma_V(t)$ & $\sigma_B(t)$ & $\sigma_R(t)$ & $\sigma_I(t)$ \\
\hline
  60.000   &  -16.938   &    0.910   &    0.201   &    0.125   &    1.496   &   -0.080   &    0.033   &    0.411   &    0.155   &    0.110   &    0.124   &    0.210  \\
   61.000   &  -16.912   &    0.901   &    0.197   &    0.088   &    1.504   &   -0.081   &    0.027   &    0.466   &    0.157   &    0.115   &    0.124   &    0.202  \\
   62.000   &  -16.887   &    0.892   &    0.193   &    0.051   &    1.515   &   -0.081   &    0.022   &    0.494   &    0.159   &    0.120   &    0.125   &    0.194  \\
   63.000   &  -16.861   &    0.883   &    0.189   &    0.013   &    1.525   &   -0.081   &    0.022   &    0.452   &    0.161   &    0.125   &    0.126   &    0.186  \\
   64.000   &  -16.835   &    0.874   &    0.185   &   -0.024   &    1.536   &   -0.081   &    0.022   &    0.425   &    0.163   &    0.130   &    0.128   &    0.178  \\
   65.000   &  -16.810   &    0.865   &    0.181   &   -0.061   &    1.547   &   -0.081   &    0.022   &    0.436   &    0.165   &    0.135   &    0.130   &    0.171  \\
   66.000   &  -16.784   &    0.856   &    0.177   &   -0.095   &    1.559   &   -0.082   &    0.022   &    0.447   &    0.167   &    0.140   &    0.133   &    0.163  \\
   67.000   &  -16.759   &    0.848   &    0.173   &   -0.127   &    1.571   &   -0.083   &    0.022   &    0.458   &    0.169   &    0.144   &    0.137   &    0.156  \\
   68.000   &  -16.733   &    0.839   &    0.169   &   -0.156   &    1.582   &   -0.084   &    0.022   &    0.470   &    0.171   &    0.148   &    0.140   &    0.150  \\
   69.000   &  -16.707   &    0.829   &    0.165   &   -0.182   &    1.594   &   -0.085   &    0.022   &    0.481   &    0.172   &    0.151   &    0.145   &    0.145  \\
   70.000   &  -16.682   &    0.821   &    0.161   &   -0.206   &    1.591   &   -0.084   &    0.022   &    0.493   &    0.174   &    0.153   &    0.149   &    0.140  \\
   71.000   &  -16.656   &    0.812   &    0.157   &   -0.226   &    1.552   &   -0.077   &    0.022   &    0.504   &    0.176   &    0.155   &    0.154   &    0.136  \\
   72.000   &  -16.631   &    0.803   &    0.153   &   -0.244   &    1.514   &   -0.070   &    0.022   &    0.516   &    0.177   &    0.156   &    0.160   &    0.133  \\
   73.000   &  -16.605   &    0.794   &    0.149   &   -0.262   &    1.491   &   -0.067   &    0.021   &    0.526   &    0.179   &    0.157   &    0.165   &    0.131  \\
   74.000   &  -16.579   &    0.785   &    0.144   &   -0.280   &    1.528   &   -0.079   &    0.021   &    0.536   &    0.180   &    0.156   &    0.171   &    0.130  \\
   75.000   &  -16.554   &    0.776   &    0.140   &   -0.298   &    1.558   &   -0.086   &    0.020   &    0.556   &    0.181   &    0.154   &    0.177   &    0.131  \\
   76.000   &  -16.528   &    0.767   &    0.136   &   -0.316   &    1.569   &   -0.081   &    0.020   &    0.588   &    0.182   &    0.152   &    0.183   &    0.133  \\
   77.000   &  -16.503   &    0.759   &    0.132   &   -0.334   &    1.580   &   -0.076   &    0.019   &    0.588   &    0.183   &    0.149   &    0.189   &    0.135  \\
   78.000   &  -16.477   &    0.750   &    0.128   &   -0.352   &    1.591   &   -0.071   &    0.019   &    0.589   &    0.184   &    0.144   &    0.195   &    0.139  \\
   79.000   &  -16.451   &    0.740   &    0.124   &   -0.370   &    1.602   &   -0.065   &    0.018   &    0.589   &    0.185   &    0.139   &    0.200   &    0.144  \\
   80.000   &  -16.426   &    0.732   &    0.120   &   -0.388   &    1.616   &   -0.060   &    0.018   &    0.588   &    0.185   &    0.133   &    0.206   &    0.150  \\
   81.000   &  -16.400   &    0.723   &    0.116   &   -0.406   &    1.626   &   -0.054   &    0.017   &    0.590   &    0.185   &    0.126   &    0.211   &    0.157  \\
   82.000   &  -16.375   &    0.714   &    0.112   &   -0.424   &    1.637   &   -0.048   &    0.017   &    0.591   &    0.185   &    0.119   &    0.216   &    0.165  \\
   83.000   &  -16.349   &    0.705   &    0.108   &   -0.442   &    1.648   &   -0.042   &    0.017   &    0.593   &    0.186   &    0.111   &    0.220   &    0.173  \\
   84.000   &  -16.323   &    0.696   &    0.104   &   -0.460   &    1.650   &   -0.036   &    0.017   &    0.595   &    0.186   &    0.102   &    0.224   &    0.182  \\
 85.000   &  -16.298   &    0.688   &    0.100   &   -0.478   &    1.615   &   -0.035   &    0.017   &    0.597   &    0.185   &    0.093   &    0.228   &    0.191  \\
   86.000   &  -16.272   &    0.678   &    0.096   &   -0.496   &    1.613   &   -0.055   &    0.018   &    0.599   &    0.185   &    0.083   &    0.231   &    0.199  \\
   87.000   &  -16.247   &    0.670   &    0.092   &   -0.514   &    1.610   &   -0.076   &    0.018   &    0.601   &    0.185   &    0.074   &    0.233   &    0.208  \\
   88.000   &  -16.221   &    0.661   &    0.088   &   -0.532   &    1.607   &   -0.097   &    0.018   &    0.603   &    0.185   &    0.064   &    0.235   &    0.216  \\
   89.000   &  -16.195   &    0.651   &    0.084   &   -0.550   &    1.606   &   -0.111   &    0.018   &    0.605   &    0.185   &    0.054   &    0.236   &    0.222  \\
   90.000   &  -16.170   &    0.643   &    0.080   &   -0.568   &    1.606   &   -0.111   &    0.018   &    0.607   &    0.184   &    0.045   &    0.237   &    0.228  \\
   91.000   &  -16.144   &    0.634   &    0.075   &   -0.586   &    1.607   &   -0.107   &    0.018   &    0.609   &    0.184   &    0.035   &    0.237   &    0.232  \\
   92.000   &  -16.119   &    0.625   &    0.071   &   -0.604   &    1.607   &   -0.107   &    0.018   &    0.609   &    0.184   &    0.027   &    0.237   &    0.235  \\
   93.000   &  -16.093   &    0.616   &    0.067   &   -0.622   &    1.607   &   -0.107   &    0.018   &    0.609   &    0.184   &    0.018   &    0.236   &    0.236  \\
   94.000   &  -16.067   &    0.607   &    0.063   &   -0.640   &    1.607   &   -0.107   &    0.018   &    0.609   &    0.184   &    0.011   &    0.235   &    0.234  \\
   95.000   &  -16.042   &    0.599   &    0.059   &   -0.658   &    1.607   &   -0.107   &    0.018   &    0.609   &    0.184   &    0.005   &    0.233   &    0.230  \\
   96.000   &  -16.016   &    0.589   &    0.055   &   -0.676   &    1.608   &   -0.107   &    0.018   &    0.609   &    0.185   &    0.000   &    0.232   &    0.224  \\
   97.000   &  -15.991   &    0.581   &    0.051   &   -0.694   &    1.608   &   -0.107   &    0.018   &    0.609   &    0.185   &    0.000   &    0.230   &    0.215  \\
   98.000   &  -15.965   &    0.572   &    0.047   &   -0.712   &    1.608   &   -0.107   &    0.018   &    0.609   &    0.186   &    0.000   &    0.228   &    0.204  \\
   99.000   &  -15.939   &    0.562   &    0.043   &   -0.730   &    1.608   &   -0.107   &    0.018   &    0.609   &    0.186   &    0.000   &    0.227   &    0.191  \\
  100.000   &  -15.914   &    0.554   &    0.039   &   -0.748   &    1.608   &   -0.107   &    0.018   &    0.609   &    0.187   &    0.000   &    0.226   &    0.176  \\
  101.000   &  -15.888   &    0.545   &    0.035   &   -0.767   &    1.608   &   -0.108   &    0.018   &    0.609   &    0.188   &    0.000   &    0.225   &    0.160  \\
  102.000   &  -15.863   &    0.536   &    0.031   &   -0.785   &    1.608   &   -0.108   &    0.018   &    0.609   &    0.189   &    0.000   &    0.224   &    0.143  \\
  103.000   &  -15.837   &    0.527   &    0.027   &   -0.803   &    1.608   &   -0.108   &    0.018   &    0.609   &    0.190   &    0.000   &    0.224   &    0.127  \\
  104.000   &  -15.811   &    0.518   &    0.023   &   -0.821   &    1.608   &   -0.108   &    0.018   &    0.609   &    0.191   &    0.000   &    0.225   &    0.112  \\
  105.000   &  -15.786   &    0.509   &    0.019   &   -0.839   &    1.608   &   -0.108   &    0.018   &    0.609   &    0.192   &    0.000   &    0.225   &    0.100  \\
  106.000   &  -15.760   &    0.500   &    0.015   &   -0.857   &    1.608   &   -0.108   &    0.018   &    0.609   &    0.193   &    0.001   &    0.227   &    0.094  \\
  107.000   &  -15.735   &    0.492   &    0.010   &   -0.875   &    1.608   &   -0.108   &    0.018   &    0.609   &    0.194   &    0.003   &    0.229   &    0.094  \\
  108.000   &  -15.709   &    0.483   &    0.006   &   -0.893   &    1.608   &   -0.108   &    0.018   &    0.609   &    0.195   &    0.005   &    0.230   &    0.105  \\
 109.000   &  -15.683   &    0.473   &    0.002   &   -0.911   &    1.608   &   -0.108   &    0.018   &    0.609   &    0.195   &    0.005   &    0.232   &    0.129  \\
  110.000   &  -15.658   &    0.465   &   -0.002   &   -0.929   &    1.608   &   -0.108   &    0.018   &    0.609   &    0.196   &    0.002   &    0.234   &    0.170  \\
  111.000   &  -15.633   &    0.457   &   -0.006   &   -0.947   &    1.608   &   -0.108   &    0.018   &    0.609   &    0.196   &   -0.001   &    0.235   &    0.210  \\
  112.000   &  -15.608   &    0.449   &   -0.010   &   -0.965   &    1.608   &   -0.108   &    0.018   &    0.609   &    0.195   &   -0.004   &    0.236   &    0.251  \\
  113.000   &  -15.583   &    0.441   &   -0.014   &   -0.983   &    1.608   &   -0.108   &    0.018   &    0.609   &    0.194   &   -0.006   &    0.238   &    0.291  \\
\hline 
\hline
\end{tabular}
\end{center}
\end{tiny}  
\end{table}

     Adding various amounts of the correction templates to the standard templates generates an empirical family of light and color curves (see figure 1).  This family of curves demonstrates some interesting relations between light curve behavior and luminosity.  The natural history of SN Ia is that intrinsically dim SN Ia rise and fall rapidly in V as compared to the leisurely rise and fall of intrinsically bright SN Ia.   These results are similar to those of Phillips (1993) with the advantage that they show the SN Ia behavior from before maximum and more than fifteen days after maximum. 

\begin{figure}
\vspace*{150mm}
\includegraphics{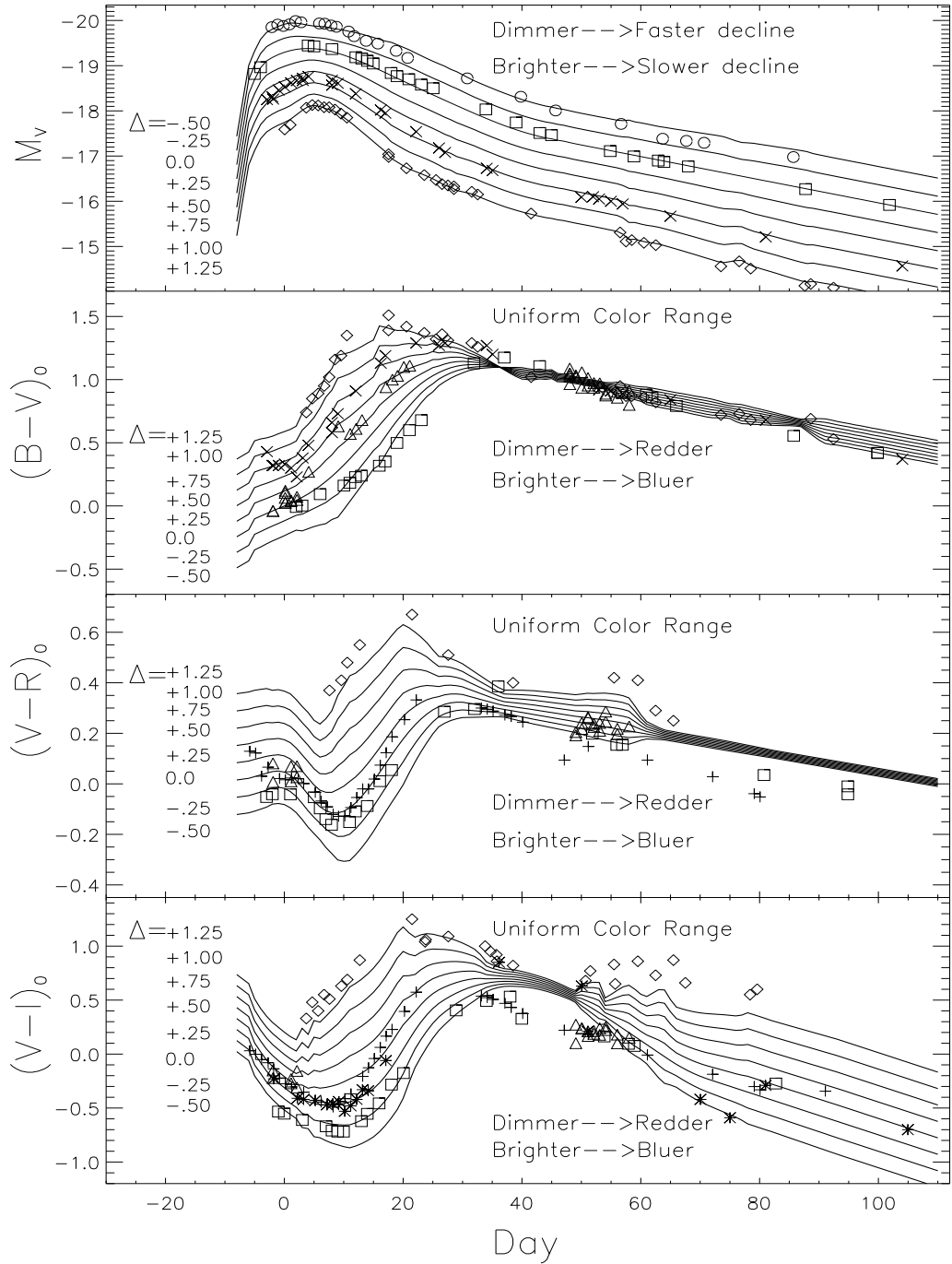}
\caption{Empirical family of SN Ia light and color curves parametrized by luminosity.  This sample of empirical V, B--V, V--R, and V--I curves is derived from the training set and depicts the entire range of light and color curve shapes and their correlation with luminosity (on the Cepheid distance scale).  This set is obtained by adding the correction templates, $R_V(t)$ or $R_{color}(t)$, multiplied by various luminosity corrections, $\Delta$, to the standard templates to make the best reconstruction of an SN Ia light and color curves.  Intrinsically dim SN Ia rise and fall faster in V and have redder colors before day 35 than intrinsically bright SN Ia.  After day 35, all SN Ia have more uniform colors.  From the multicolor light curve shape (MLCS) method we estimate the luminosity and extinction by dust independently from the distance to measure the extinction-free distance.  Data shown as reconstructed, 91T=$\circ$, 94ae=$\Box$, 86G=X,91bg=$\diamond$, 92A=+, 80N=$\triangle$}
\end{figure}
\eject

  An interesting new result is the quantitative relation between SN Ia luminosity and color.  Supernova luminosities correlate with their intrinsic colors at early times (see also Lira, P. 1995).  Before day thirty-five, the dim SN Ia  are red and the bright ones are blue.   For example, at maximum, a decrease of 0.10 magnitudes in visual absolute luminosity corresponds to an increase in color (toward the red) of $0.05$, $0.03$, and $0.04$ magnitudes in B$-$V, V$-$R, and V$-$I, respectively.   Presuming that all SN Ia have a uniform color at maximum (as has often been done) is a poor assumption that can lead to incorrect predictions as we show elsewhere (Riess, Press, Kirshner 1996b).   In particular, the relation between intrinsic color and luminosity helps explain how absorption ``corrections'' have led to {\it increased} dispersion in Hubble diagrams and to unphysical properties derived for dust in distant galaxies (Branch \& Tammann 1992, Capaccioli et al 1990, Joeveer 1982, Tammann 1987, Sandage et al 1993).   For comparison, the color variations at maximum due to reddening for 0.10 magnitudes of visual extinction are expected to be $0.03, 0.03$, and $0.05$ in B$-$V, V$-$R, and V$-$I.   Although the sense of the color changes from absorption is the same as from intrinsic variation, the values are not.

  van den Bergh (1995) noted that the relation between intrinsic B--V color and luminosity predicted by the mean behavior of H\"{o}flich and Khokhlov's (1996) theoretical models coincidentally agrees with the standard reddening law.  He takes advantage of this agreement to define a reddening-free luminosity which employs a single measurement of B--V color to account for both the luminosity variation intrinsic to the supernova and that which results from absorption by dust.  Our empirical relation between luminosity and color shows that dust and intrinsic luminosity variation do not cause exactly the same change in color.  This difference may cause the increase in dispersion around the Hubble line observed after van den Bergh's prescription is applied (Riess, Press, Kirshner 1996b).

  More than thirty-five days past maximum light, all supernovae exhibit nearly uniform colors.   Entries in the correction templates for day thirty-five give differences of $0.00$, $0.01$, and $0.01$ magnitudes in B$-$V, V$-$R, and V$-$I color for a 0.10 magnitude change in visual absolute luminosity.  Detailed theoretical modeling of SN Ia show a similar relation between supernova color and luminosity (H\"{o}flich \& Khokhlov 1996).

   An alternate view of the photometric differences between intrinsically bright and dim SN Ia is presented in figure 2.  The family of absolute B,V,R, and I light curves show recognizable morphological variations.  The B light curve family is similar in behavior to the V family;  dim SN Ia rise and fall more rapidly in B and V than bright SN Ia.  In R the brighter SN Ia have a ``shoulder'' $\sim$ 25 days after B maximum.  For dimmer SN Ia, this shoulder is less pronounced and disappears completely for the most underluminous objects.  In the I band, the brighter SN Ia have two maxima.  The first occurs quite early, $\sim$ 5 days before the B band maximum.  The second, broad maximum is at $\sim$ 30 days after B maximum.  As the luminosity of the SN Ia decreases, a number of changes in the I band light curve are apparent: the first maximum is later and broader while the second maximum is dimmer and occurs earlier.  For the most underluminous SN Ia, the two maxima merge into one maximum which is broad and occurs $\sim$ 5 days after B maximum.

\begin{figure}
\vspace*{150mm}
\includegraphics{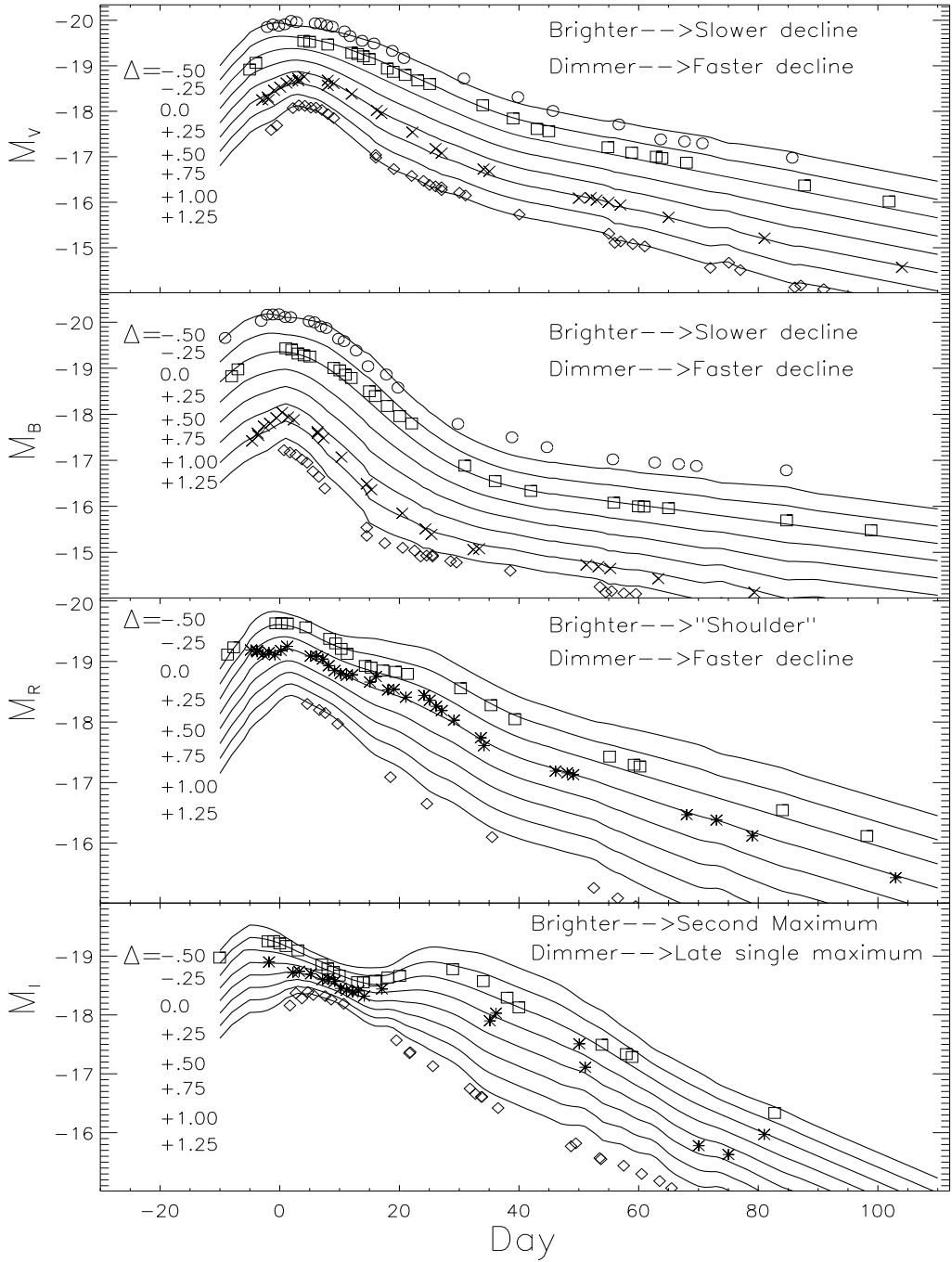}
\caption{Empirical family of SN Ia BVRI light curves parameterized by luminosity.  This family of light curves is derived in the same way as the families in figure 1 and shows the differences in photometric behavior for bright and dim SN Ia.  Intrinsically dim SN Ia rise and fall faster in B and V than intrinsically bright SN Ia.  For the R light curve, a ``shoulder'' occurs $\sim$ 25 days after B maximum in the bright SN Ia.  This shoulder is weaker for dimmer SN Ia and is absent for the most underluminous ones.  In the I band, the bright SN Ia have two maxima; one early ($\sim$ 5 days before B maximum) and one later ($\sim$ 30 days after B maximum).  As the luminosity of the SN Ia decreases the first maximum occurs later and is broader while the second maximum is dimmer and occurs earlier.  For the most underluminous SN Ia, the two maxima merge into one maximum which is broad and occurs $\sim$ 5 days after B maximum.  Data shown as reconstructed, 91T=$\circ$, 94ae=$\Box$, 86G=X,91bg=$\diamond$, 92A=+, 80N=$\triangle$}
\end{figure}
\eject

  The light and color curve reconstructions in figure 1 provide a powerful means to measure the extinction-corrected distance at every phase of supernova observation.  SN Ia appear dim because they are distant, obscured by dust, or intrinsically dim.  We can distinguish between these possibilities by using the light and color curve shapes (which determine the intrinsic luminosity and color) and measuring the observed offsets (which determine the extinction-corrected distance).  
  
\section{Extinction-corrected Distances from Multicolor Light Curve Shapes}  

     Figure 1 provides a ready guide for measuring extinction-corrected distances.  Given BVRI light curve photometry, we seek the best set of curves for a fixed value of $\Delta$ which minimizes the $\chi^2$ between model and data.  The offsets between the model and the data provide the best estimates of $\mu$ and $A_V$.  Rather than searching the $\chi^2$ parameter space for a solution we take advantage of the linearity of our model for an immediate solution.

     Referring to the matrix form of the model in equation (9), we use the following definitions; {\bf y} is the column of apparent magnitude measurements,  {\bf s} is the column of standard templates, {\bf L} is the three column matrix with correction templates and offsets, and {\bf q} is the three element column of parameters.  With these definitions, we rewrite $\chi^2$ in equation (10) as 
\bq \chi^2=({\bf y}-{\bf s}-{\bf Lq})^T{\bf C}^{-1}({\bf y}-{\bf s}-{\bf Lq}).\eq The analytical minimization of $\chi^2$ with respect to the column of free parameters {\bf q} gives

\bq {\bf q_{{\it best}}}=[ \mu \ \ A_V \ \ \Delta]_{best}^T= ({\bf L}^T{\bf C}^{-1}{\bf L})^{-1}{\bf L}^T{\bf C}^{-1}[{\bf y}-{\bf s}]  \eq   Equation (14) {\it simultaneously} measures the distance and the extinction using all the available light curve observations, reducing the reliance on any particular time of the supernova light curve.  According to our definitions in equation (1) or (5), $\mu$ is the apparent distance modulus {\it uncorrected for extinction}.  The extinction-corrected MLCS distance is given by $\mu - A_V$.  The ``standard candle'' distance, i.e. without any correction for light curve shape (luminosity) or reddening, is given by $\mu + \Delta$.  This is the distance derived by correctly fitting the shape of the visual light curve to find the peak, ignoring the color excesses, and assigning the SN Ia a standard luminosity.   The standard errors for the parameters in ${\bf q}_{best}$ are given by the covariance matrix, $({\bf L}^T{\bf C}^{-1}{\bf L})^{-1}$.  These errors are the {\it fitting} errors which reflect the uncertainty in locating a light curve's best placement in figure 1, due to the presence of noise in either the training set light curves or in the independent light curve we are trying to fit.  

This approach, until now, treats the correction templates derived from our training set as if they were perfect.  This is certainly not the case since the independent distance and A$_V$ estimates for the objects that make up the training set objects are themselves not perfect.  The same type of uncertainty is more easily seen in the Phillips (1993) relation where the light curve decline in the first fifteen days after maximum is correlated against the luminosity derived from independent distance and A$_V$ estimates.  The uncertainty in the slope of this relation must be considered when it is used.  Fortunately, our correction templates are well constrained by the accuracy of the distance and A$_V$ estimates combined with the size of the training set.  The external source of error on the parameters  can be found by varying the training set distances and A$_V$ estimates in a Monte Carlo simulation to determine the effect on the fitted parameters.  As expected, the external distance error increases linearly with the light curve's luminosity correction, $\Delta$, and is well described by $\sigma=0.055\Delta$ mag.  So for a supernova whose luminosity correction $\Delta=0.30$ mag, our external distance error would amount to less than 0.02 mag.  This error is negligible for the majority of observed supernovae whose typical $\vert \Delta \vert \leq 0.50$ mag results in $\sigma \leq 0.03$ mag.  The external source of error will decrease as the square root of the training set size, which we will be able to expand in the future.

    The time of maximum for each supernova is a non-linear parameter which cannot be solved for analytically in this scheme.  It requires an outer iteration to minimize $\chi^2$ in equation (13).  For our well-observed light curves, the uncertainty in the time of maximum, typically well under one day, has a negligible effect on the parameter errors, with the median {\it error} increasing by only 10\% if the time of maximum is varied $\pm$ 1 day.   For poorly observed light curves whose observations begin $\sim 10$ days after maximum, the uncertainty in the time of maximum increases substantially as does its effect on the parameter errors.   We discard all light curves beginning more than 10 days after maximum to avoid SN Ia with large uncertainties while maintaining a useful number of objects.  When the set of usable SN Ia has grown sufficiently, the precision of this distance indicator might be improved by imposing an even stricter requirement for the time of the first observation.

\section{Constructing the Correlation Matrix}

The correlation matrix, {\bf C}, used in equations (13) and (14) to determine the best parameters and $\chi^2$ of the light curve fit, is the sum of two parts, {\bf C=S+N}.  The noise correlation matrix,  {\bf N}, is the correlation matrix for the measurement errors.  The signal correlation matrix, {\bf S}, is the correlation matrix which, {\it in the absence of measurement error}, estimates the expected deviations of the light curves from our model.  It is the {\bf S} matrix which allows us to use our model despite its known shortcomings.  The {\bf N} matrix is supplied by the conscientious observer.  

The {\bf S} matrix has two parts, diagonal and off-diagonal entries, which, in principle, are estimated from our training set.  The diagonal entries are estimates of the expected deviation of the light curve from the model.  We determine the entries along the diagonal of each photometric band's block of {\bf S} by measuring the dispersion of each training set member's $R_V$ and $R_{color}$ around the ensemble average $R_V$ and $R_{color}$ given in equations (11) and (12) minus each light curve's contribution from measurement error.  The result gives, as a function of time, the expected errors of noise-corrected data around the best fit model.  We have plotted this dispersion around the best fit model reconstruction in figure 3 using the standard templates ($\Delta=0$) as an example.  The resulting ``grey snakes'' comprise 1 sigma confidence regions which, when plotted over a best model fit, would be expected to contain 68\% of the data points.  The square root of the diagonal entries of the ${\bf S_{V,B,R,I}}$ matrix are provided in table 2.

\begin{figure}
\vspace*{150mm}
\includegraphics{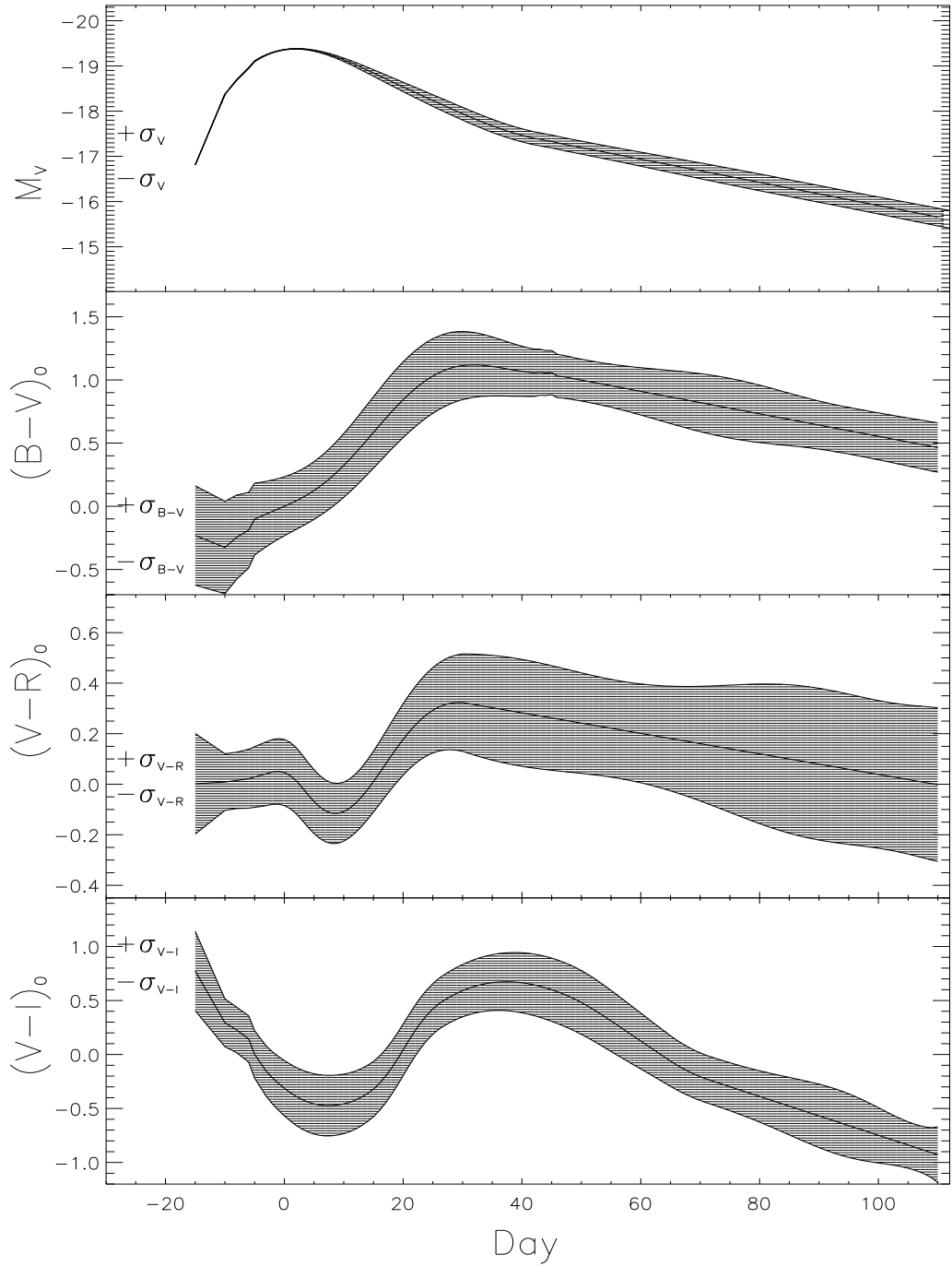}
\caption{Dispersion of noise-free light and color curves around the best fit model.    The ``grey snakes'' comprise 1 sigma confidence regions which, when plotted over a best fit model fit (here chosen as the $\Delta=0$ standard templates), are expected to contain 68\% of noise-free data points.  The square of these functions would be the diagonal entries of the signal correlation matrix ({\bf S}), which, added to the noise correlation matrix ({\bf N}), would give the model correlation matrix ({\bf C}).   To compensate for our inability to determine either data covariance (off diagonal elements) or higher order correlations in luminosity we rescale each of these functions as described in \S 4.}
\end{figure}
\eject  

The off-diagonal entries of the {\bf S} matrix provide estimates of the correlation between model residuals. These correlations are likely to be considerable.  There are many more elements to consider in determining the point-to-point correlations than 
we had for the previous autocorrelations.  While initially we had only to estimate the expected model residuals for a given band on a given day of the light curve, the off-diagonal entries require much more information.  We would need to estimate the amount of covariance between observations on different days in the same band, in different bands on the same day, and in different bands on different days (In addition, we would have to remove the contribution from measurement covariances which observers generally do not supply!).  Unfortunately, our sparse training set is currently inadequate for quantifying these covariances.  A simple two-point correlation function suggests that these covariances are most important in B, R, and I but does not provide enough information to approximate them adequately.  This is the same compromise we faced in choosing a simple linear model over one with higher order terms.   A detailed description of  how our data deviates from a linear model requires the same information as would be required to establish a more detailed model for our data.  Either improvement requires a larger training set.  We echo our minimalist approach to our model with a minimalist approach to the {\bf S} matrix.  We employ a diagonal {\bf S} matrix (with the diagonals determined as above) with compensation for the possibility that some photometric bands are more adequately modeled with our linear model than others.  We increase the diagonal elements of the signal correlation matrix enough to compensate for our inability to estimate its off-diagonal terms.   A simple rescaling of the entire signal correlation matrix would be less effective because some photometric bands of data (matrix blocks) have more model covariance (i.e. non-linear behavior) than others.      

We seek to weight each band's data in the correlation matrix by its ability to predict the parameter in common, $\Delta$. This approach recognizes that in a linear model some bands may be better than others at estimating the parameter $\Delta$ and weights each light curve accordingly.  To determine each band's weight, we allow their weights in the correlation matrix to vary and maximize the log-likelihood function for the determination of $\Delta$

 \bq {\cal L} \equiv -{1 \over 2 }(\chi_{\Delta}^2 -\sum_{i=1}^n ln{ 1 \over   \sigma_{\Delta_i}^2} ). \eq  Maximizing ${\cal L}$ is the desired way to determine parameters of the correlation matrix where the conventional approach of minimizing $\chi^2$ would necessarily drive the weights and $\chi^2$ to zero (Rybicki and Kleyna, 1994).  A simple exercise shows that maximizing ${\cal L}$ with respect to the weights drives the residuals to the smallest values they can have while still maintaining a $\chi^2 $ per degree of freedom near unity.  By maximizing ${\cal L}$, we optimize our data's ability to predict the parameter, $\Delta$, while simultaneously requiring that the estimated error in $\Delta$ is reasonable. Given more training set data, we could estimate the entire correlation matrix by maximizing  ${\cal L}$. With our current limited training set, we will use our previous prescription to parametrize the diagonal  blocks of the correlation matrix that correspond to each photometric filter and use ${\cal L}$ to determine the relative weights of those blocks.  

 Here  ${\cal L}$ takes the form
\bq   {\cal L}=-{1 \over 2 }(\sum_{i=1}^n { (\Delta_{MLCS} - \Delta_{ind})^2 \over   \sigma_{\Delta_{MLCS}}^2+\sigma_{\Delta_{ind}}^2}-\sum_{i=1}^n ln({ 1 \over   \sigma_{\Delta_{MLCS}}^2+\sigma_{\Delta_{ind}}^2})) \eq where the subscripts {\it ind} and {\it MLCS} denote the value and uncertainty of $\Delta$ as determined by independent methods and MLCS respectively.  Optimal weighting of the B,V,R, and I data minimizes the difference between the independently determined $\Delta$'s (in table 1) and the MLCS predicted $\Delta$'s which are a function of the weights in the {\bf C} matrix.  Using equation (16) and a downhill simplex method (Press et al 1992) to find its maximum, we have determined the relative weights of each band.  The diagonal  ${\bf C_{V,B,R,I}}$ matrix needed for the MLCS fit comes from adding the square of the model errors in table 2 to the square of the observers' photometry errors, then multiplying the four diagonal blocks of ${\bf C_{V,B,R,I}}$ by $0.37, 11.45, 8.85, 5.52$.   These values, determined by maximizing equation (16), indicate that the V band light curves are significantly better predictors of $\Delta$ than B,R, or I data within the framework of our linear model.  Specifically, a V band observation contributes 5.6, 4.9, or 3.9 times as much as a B, R, or I band observation towards determining the luminosity correction for a SN Ia.  This result is not astonishing since it is well established that light curves in B,R, and I with differing luminosity can cross each other (Suntzeff, 1993).  A crossing point of such light curves implies the same light or color curve shape for SN Ia with different $\Delta$'s.  Such behavior found in B, R, and I diminishes their predictive power in our linear model, but the independence of their $A_V$ measurements provide useful estimates of extinction.  

We have assumed the correlation matrix for V,B,R, and I data is diagonal, but the correlation matrix for V,B-V,V-R, and V-I is certainly not diagonal since an observation of $m_V$ and $m_{B-V}$ made at the same time are anti-correlated with covariance -$\sigma^2_V$.   Formation of the correlation matrix for V,B-V,V-R, and V-I data, i.e. ${\bf C_{V,B-V,V-R,V-I}}$, may be done directly with care to include the covariance terms of $\pm \sigma^2_V$ or by means of a simple rotation.  This involves writing down the rotation matrix, {\bf A}, which satisfies         
\bq {\bf A} \pmatrix{V \cr B-V \cr V-R \cr V-I} = \pmatrix{V \cr B \cr R \cr I}. \eq  and we readily identify \bq {\bf C^{-1}_{V,B-V,V-R,R-I}}= {\bf A^T} {\bf C^{-1}_{V,B,R,I}} {\bf A}. \eq  Deriving the desired correlation matrix from equation (18) has the advantage of requiring the much simpler {\bf A} and diagonal ${\bf C^{-1}_{V,B,R,I}}$ matrices.
     
Meaningful model parameter errors can only come from models which fit the data within statistical expectations.  We require that both the V light curve model and the color curve models give a reduced (per degree of freedom) $\chi^2$ of 1.  To define the confidence region for the distance parameters, we use the covariance matrix of the fit, $ ({\bf L}^T{\bf C}^{-1}{\bf L})^{-1}$, multiplied by the {\it reduced} $\chi^2$ of the light or color curve fits which measures the parameter of interest.  This, in effect, is renormalization of the {\bf C} matrix, now done on a supernova by supernova basis.  This renormalization is not generally a large factor and would presumably become unnecesary with a better non-diagonal model for {\bf C}.

   In the case of the distance modulus error, derived from the visual band data, the error is the (1,1) entry of the covariance matrix multiplied by the reduced $\chi^2$ of the V light curve fit:
 \bq \sigma^2_{\mu}= ({\bf L}^T{\bf C}^{-1}{\bf L})^{-1}_{\ 1 \ 1}  \chi^2_{\nu}(V). \eq
Similarly, the extinction error, as derived from the B--V, V--R, V--I, is the (2,2) entry of the covariance matrix multiplied by the reduced $\chi^2$ of the color curves' fit:
\bq \sigma^2_{A_V}= ({\bf L}^T{\bf C}^{-1}{\bf L})^{-1}_{\ 2 \ 2}  \chi^2_{\nu}(B-V,V-R,V-I). \eq
The extinction-corrected distance is given by $\mu-A_V$ and its variance is the sum of equations (19) and (20) minus twice the covariance of the estimates of $\mu$ and $A_V$:
\bq  \sigma^2_{\mu-A_V}=\sigma^2_{\mu} + \sigma^2_{A_V} -2 ({\bf L}^T{\bf C}^{-1}{\bf L})^{-1}_{\ 1 \ 2} \sqrt{\chi^2_{\nu}(V) \chi^2_{\nu}(B-V,V-R,V-I)}. \eq

The extinction-corrected distance error of equation (21) is the previously mentioned fitting error.  For a particular SN Ia, its size depends on light curve sampling, measurement errors and light curve shape (see \S 6).  These errors provide useful individual estimates of distance uncertainty.

\section {Formalized Truncation of $A_V$}
    What is the best way to estimate the absorption by dust given a measurement of excess color?  We have well-founded {\it a priori} knowledge that dust scatters or absorbs light but does not amplify it so the true value of $A_V$ must lie within the range \bq 0\leq A_V \leq (\mu+M_V-m_{lim}) \eq where $m_{lim}$ is a detection limit.  Since we measure $A_V$ by dust's reddening effect we say, {\it a priori}, that dust cannot ``blue-en'' or brighten an SN Ia.  We could use this knowledge to improve our estimate of $A_V$ by truncating any measurement of $A_V$ found to be less than zero.  Simple truncation carries the disadvantage of improperly treating our useful estimate of $\sigma_{A_V}$ from equation (20).  What is the best way to use our estimate of $A_V$ and its error together with prior knowledge that $A_V$ cannot be less than zero?  A straightforward Bayesian calculation provides the solution.
  
Suppose we were to make an estimate of $A_V$ with value
$\hat{a}$ and normal error $\sigma_{\hat{a}}$.  Further, we have some knowledge of the distribution of the {\it observed} $A_V$, $p(A_V)$.  Using Bayes's theorem

\bq p(A_V \vert \hat{a},\sigma_{\hat{a}})={p(\hat{a} \vert A_V,\sigma_{\hat{a}}) p(A_V) \over p(\hat{a})} 
= { e^{-{ (A_V-\hat{a})^2 \over 2\sigma^2_{\hat{a}} }}p(A_V) \over \int_0^{\infty} p(A_V)\: e^{-{ (A_V-\hat{a})^2 \over 2\sigma^2_{\hat{a}} }}\,dA_V} \eq
 
\begin{figure}
\vspace*{150mm}
\includegraphics{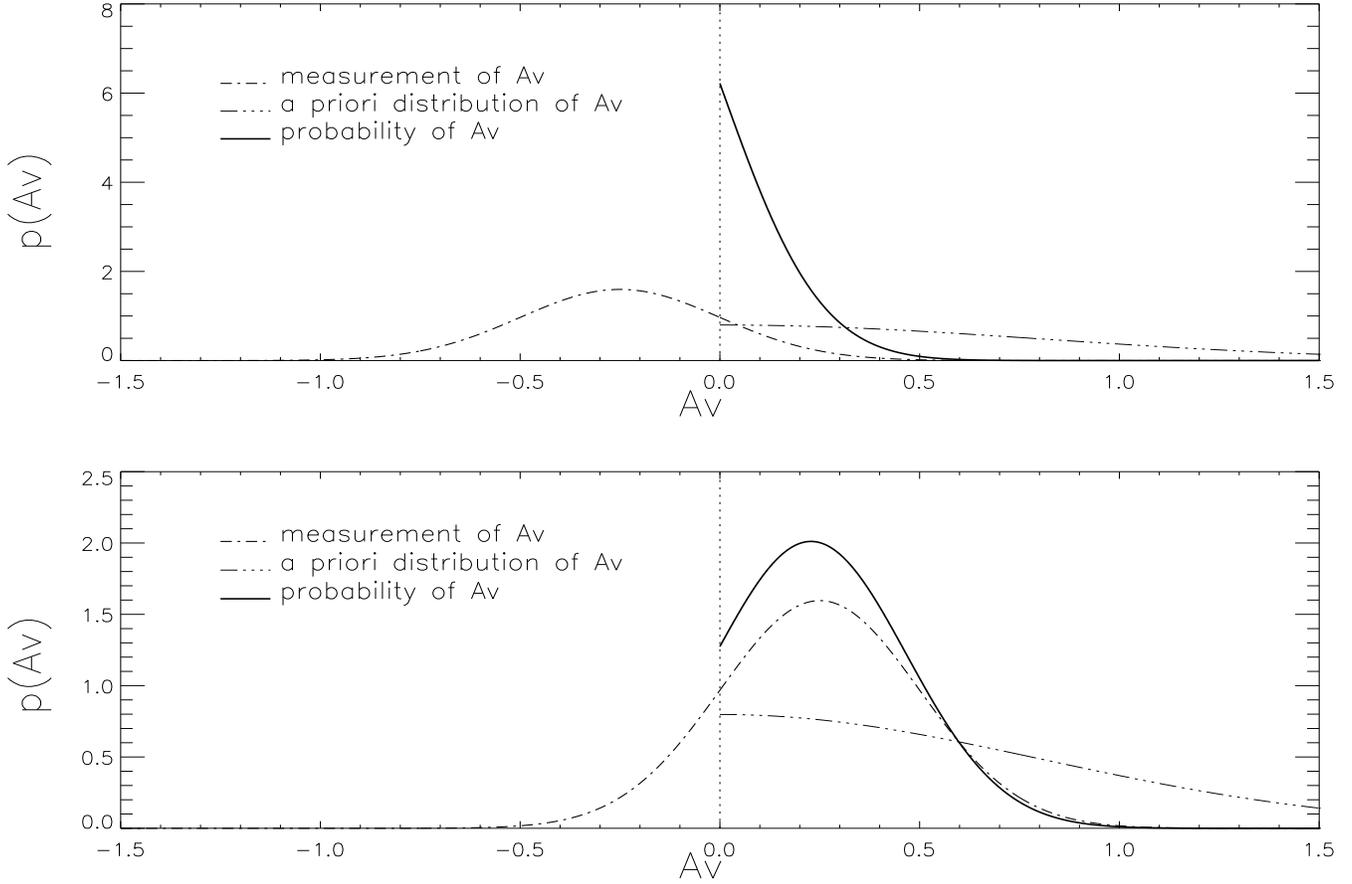}
\caption{Bayesian estimation of visual extinction by dust.  By combining our MLCS measurement of visual extinction, $A_V$, and its error with the {\it a priori} knowledge that $A_V$ is positive we can estimate the likelihood and distribution of the {\it true} value for $A_V$.  The top panel combines a {\it negative} estimate for $A_V$ (=$-0.25 \pm 0.20$) with a gaussian wing ($\sigma=1$) {\it a priori} distribution to yield an estimate for $A_V$(=$0.0 \pm 0.08$).  The bottom panel combines a {\it positive} estimate for $A_V$ (=$0.25 \pm 0.20$) with the same {\it a priori} distribution to give an estimate for $A_V$(=$0.24 \pm 0.16$).}
\end{figure}
\eject 

 This ``Bayesian filter'' provides a probability distribution for the {\it true} $A_V$ from which we can obtain a best estimate of $A_V$ and its error.  A minimalist approach to $p(A_V)$ would be to assume it is constant over the range in equation (22).  This formalized truncation is too conservative and unrealistic and we can do better.  Since supernovae with very large values of $A_V$ are less likely to be part of a sample of detected supernovae, we suggest a reasonable form for the {\it observed} $p(A_V)$ is a one-sided gaussian which has a maximum at $A_V$=0 and declines for large $A_V$ (see figure 4).   We have chosen $\sigma$=1 magnitude of $A_V$ for our $p(A_V)$, but we show in \S 7 that our results are insensitive to the particular value used for $\sigma$. 

\section {Comparing Distances} 

   The above completes our development of an algorithm to measure extinction-corrected distances with multicolor light curve shapes.  There are, however, a few complications to consider in the practice of measuring distances to supernovae with light curve shapes.
        
    The ``K correction'' (Oke and Sandage 1968, Humason, Mayall, and Sandage 1956) corrects for the effects of redshift on the measured flux through a filter of fixed spectral response.  These corrections can be approximated by measuring the effect of redshifting the spectra of SN Ia supernovae for different redshifts and phases.  We can include K corrections in our templates given the redshift, or include them in our measurements given the redshift, using an assumed time of maximum in the outer iteration.  For B and V band data we have used the K corrections of Hamuy et al (1993b) and for R and I we have calculated our own (see Filippenko et al 1996). The light curves are also affected by time dilation so we contract the light curve by (1+z) to return it to the rest frame (Leibundgut et al. 1996, Goldhaber et al 1996).          

      Using equation (14) we now measure the distance related parameters for a set of well-observed SN Ia assuming that they share the same behavior as our training set.  By applying the method as developed for the training set to this independent sample, we compute distances for each supernova, and construct a Hubble diagram.  Analysis of this diagram shows that the MLCS approach gives better precision than the standard candle method.  We will also compare the results with a sample selected by the same criteria used by Tammann and Sandage (1996) to see whether MLCS improves the utility of a ``normal'' set of SN Ia.   We restrict our attention to light curves obtained on a modern photometric system where the light curve begins within ten days of maximum light (as determined by our fit).   Tests on the training set have shown that in order for a light curve to contain luminosity information in its shape the first observation must be within ten days of maximum.  Every one of the SN Ia in our samples was recorded on digital images and has accurate subtraction of the host galaxy background to insure the light curve shapes are free from systematic errors (Boisseau and Wheeler 1991).

\begin{table}[bp]    
\begin{center}    
\caption{SN Ia Parameters} \vspace{0.4cm}
\begin{tabular}{lcccccc} \hline \hline 
  {\em SN Ia}     &   $ \log v (km s^{-1})$       & $\mu-A_V$   & $\sigma_{\mu-A_V}$  & $\Delta$ & $A_V$ & galactic $A_V$ \\
\hline   
1992bo  &     3.734  &     34.59  &      0.07  &      0.51  &      0.00  &      0.00\\
1992bc  &     3.779  &     34.75  &      0.05  &     -0.23  &      0.00  &      0.00\\
1992K  &     3.521  &     33.53  &      0.15  &      1.25  &      0.01  &      0.23\\
1992aq  &     4.481  &     38.27  &      0.06  &      0.35  &      0.00  &      0.00\\
1992ae  &     4.350  &     37.79  &      0.09  &     -0.05  &      0.00  &      0.04\\
1992P  &     3.896  &     35.50  &      0.08  &     -0.20  &      0.11  &      0.00\\
1992J  &     4.140  &     36.75  &      0.14  &      0.35  &      0.18  &      0.15\\
1991U  &     3.991  &     35.70  &      0.17  &     -0.41  &      0.75  &      0.20\\
1991ag  &     3.613  &     34.15  &      0.08  &     -0.27  &      0.01  &      0.13\\
1990af  &     4.178  &     36.84  &      0.06  &      0.27  &      0.00  &      0.05\\
1992G  &     3.299  &     32.22  &      0.11  &      0.02  &      0.45  &      0.04\\
1991m  &     3.389  &     32.96  &      0.10  &      0.73  &      0.00  &      0.01\\
1993ae  &     3.709  &     34.52  &      0.11  &      0.38  &      0.00  &      0.10\\
1994M  &     3.859  &     35.32  &      0.09  &      0.21  &      0.00  &      0.00\\
1994S  &     3.685  &     34.24  &      0.05  &     -0.13  &      0.00  &      0.00\\
1994T  &     4.030  &     36.16  &      0.10  &      0.33  &      0.00  &      0.00\\
1994Q  &     3.938  &     35.86  &      0.13  &     -0.28  &      0.25  &      0.05\\
1993ac  &     4.170  &     36.93  &      0.22  &      0.34  &      0.20  &      0.45\\
1995D  &     3.398  &     32.76  &      0.06  &     -0.28  &      0.21  &      0.00\\
1995E  &     3.547  &     33.76  &      0.06  &     -0.17  &      1.86  &      0.00\\
\hline 
\hline
\end{tabular}
\end{center}
\end{table}

Our independent set of twenty supernovae contains ten objects from the Calan/Tololo survey (Hamuy et al. 1993a, 1994, 1995, 1996; Maza et al. 1994), two from the literature  (Ford et al. 1993) and eight from our own work (Riess et al. 1996).  Table 3 contains SN Ia MLCS parameters.  Host galaxy redshifts (column 2) are in the CMB rest frame.  The heliocentric redshifts (Riess et al 1996, Ford et al 1993, Hamuy 1995) are transformed to the Local Group rest frame by the addition of (--30,297,--27) km s$^{-1}$ in Galactic Cartesian coordinates  (de Vaucouleurs et al 1991, Lynden-Bell \& Lahav 1988).  These Local Group redshifts are transformed to the CMB rest frame with the addition of (10,--542,300) km s$^{-1}$ (Smoot et al. 1992).  For SN 1993ae we have used the redshift of Abell 194 (Chapman 1988) of which the host galaxy is a member.  For the three SN Ia with cz $\leq$ 3000 km s$^{-1}$ (SN 1991M, SN 1992G, and SN 1995D) we have corrected their redshifts for their likely infall towards Virgo (Schmidt, Kirshner, \& Eastman 1992).  

The Galactic extinction measures (column 7) which we use for comparison to our own are from Burstein \& Heiles (1982).  The values of $\mu-A_V$ and $\sigma_{\mu-A_V}$  in columns 3 and 4 are the MLCS extinction-corrected distances and the fitting errors with the former placed onto the Cepheid variable distance scale as described in \S 7 and table 5.  The line-of-sight extinction estimate, $A_V$, is listed in column 6 and adding it to the values of $\mu-A_V$ gives the distance estimate without any correction for absorption.  Adding these values of $\mu$ to the luminosity correction, $\Delta$, in column 5 gives a distance estimate without correction for either the luminosity-light curve relation or absorption.  

   In figure 5 we show the multicolor light curve shape reconstruction for three SN Ia (Riess et al 1996) from our independent sample spanning the range of data quality and distance error.  For SN 1993ac we have 23 observations beginning shortly after maximum light resulting in an extinction-corrected distance error of 0.20 mag.  The light and color curves of SN 1994Q contain 42 observations beginning shortly after maximum and give an extinction-corrected distance error of 0.13 mag.  SN 1995D has one of the best sampled light and color curves with 107 observations beginning before maximum light and yielding an extinction-corrected distance error of 0.06 mag.  In general, the size of our predicted extinction-corrected distance error depends on the number of observations, the noise in the observations, and whether the SN Ia was first observed before or after maximum light.  Half of our independent sample of twenty SN Ia were observed at or before maximum light.

\begin{figure}
\vspace*{150mm}
\includegraphics{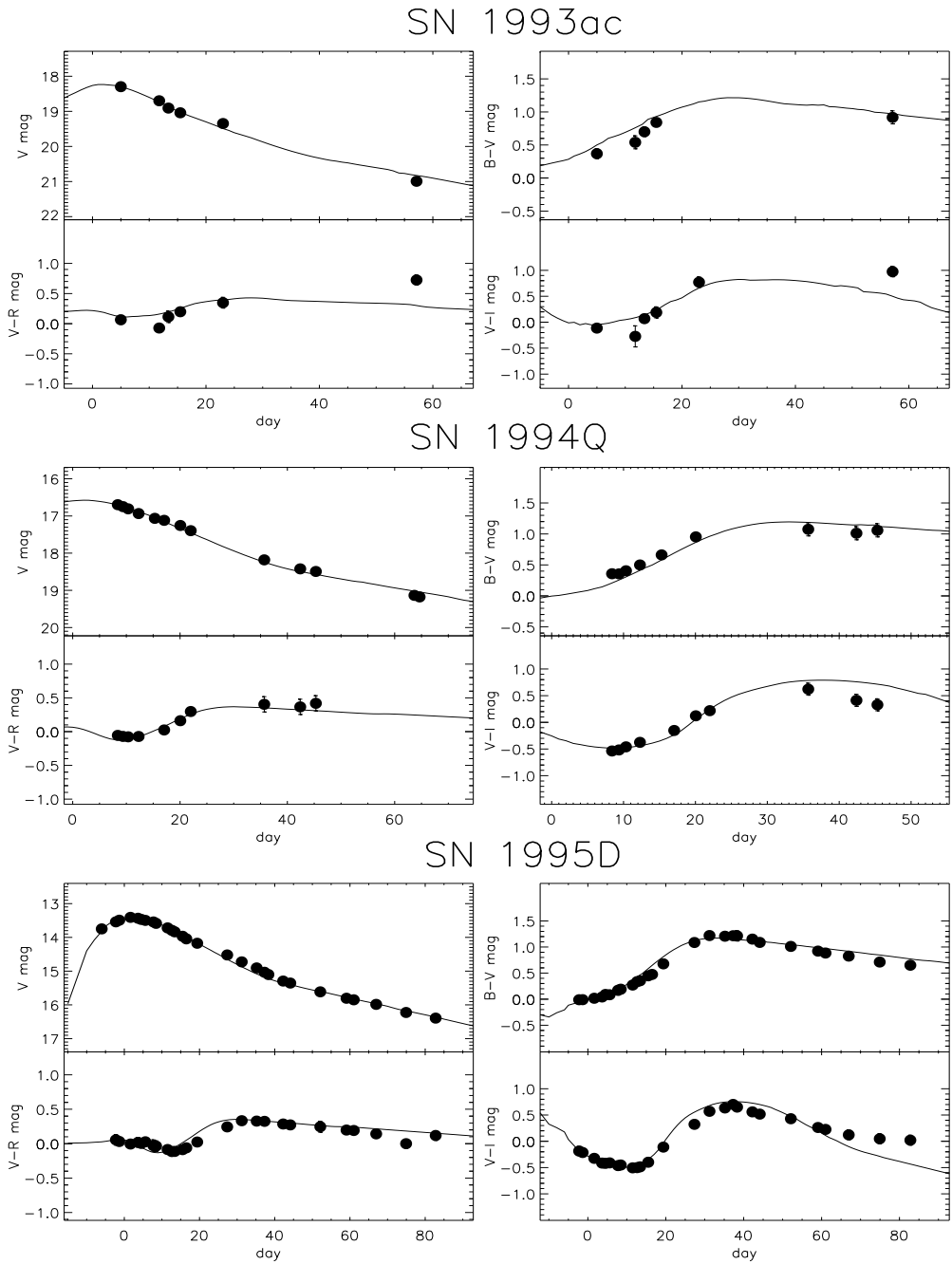}
\caption{Three SN Ia MLCS fits spanning the range of data quality in the independent sample of twenty SN Ia.  SN 1993ac has 23 noisy observations beginning shortly after maximum light resulting in a distance error of 0.20 mag.  SN 1994Q has 42 observations with low noise beginning after maximum and has a distance error of 0.13 mag.  SN 1995D has 107 observations with little noise beginning before maximum resulting in a distance error of 0.06 mag.  Photometry is from Riess et al (1996).}
\end{figure}
\eject

 First, we will use the Hubble diagram as an analytical tool, without reference to a distance scale calibration, to determine the precision of our method.    Figure 6 shows, for comparison,  two Hubble diagrams for this independent set of supernovae.  In figure 6a we have fit the best light curve shape to each supernova to estimate the distance {\it without any correction for intrinsic luminosity variation or extinction}.   In figure 6b we have plotted the MLCS extinction-corrected distances to each SN Ia which accounts for intrinsic luminosity variation and extinction (see table 4).  The reduction in dispersion is dramatic.  The improvement  in distance precision comes from deriving the correlation between luminosity, and light and color curve shape from our training set of SN Ia, and then applying these relations to the independent sample.  Because the training set and the independent set have no overlapping members, the reduction in dispersion is a powerful demonstration of the effectiveness of the MLCS method.

Table 4 compares the distance estimates for different assumptions by measuring the dispersion and $\chi^2$ on the Hubble diagram.   In each case we have made a custom reconstruction of the light and color curves.    The first row is the ``standard candle'' assumption for which we disregard the light curve shape-luminosity correction, $\Delta$, by adding it back to the distance, $\mu+\Delta$, and make no correction for extinction.   Next,  we use the MLCS distance modulus, $\mu$ which includes the luminosity information, but make no allowance for absorption.  Following this we include a correction for only the Galactic component of extinction by using Burstein \& Heiles (1982) absorption measures.  Finally, we use the full MLCS method to estimate the extinction-corrected distance.  We compare the different methods for three different subsamples: all twenty SN Ia, all SN Ia minus the single most highly reddened object (SN 1995E), and with a color cut (Vaughan et al 1995, Tamman \& Sandage 1995, Hamuy et al 1995) which discards objects outside the color range $-0.25 \leq (B-V)_{max} \leq 0.25$, leaving eighteen objects.   

\begin{figure}
\vspace*{150mm}
\includegraphics{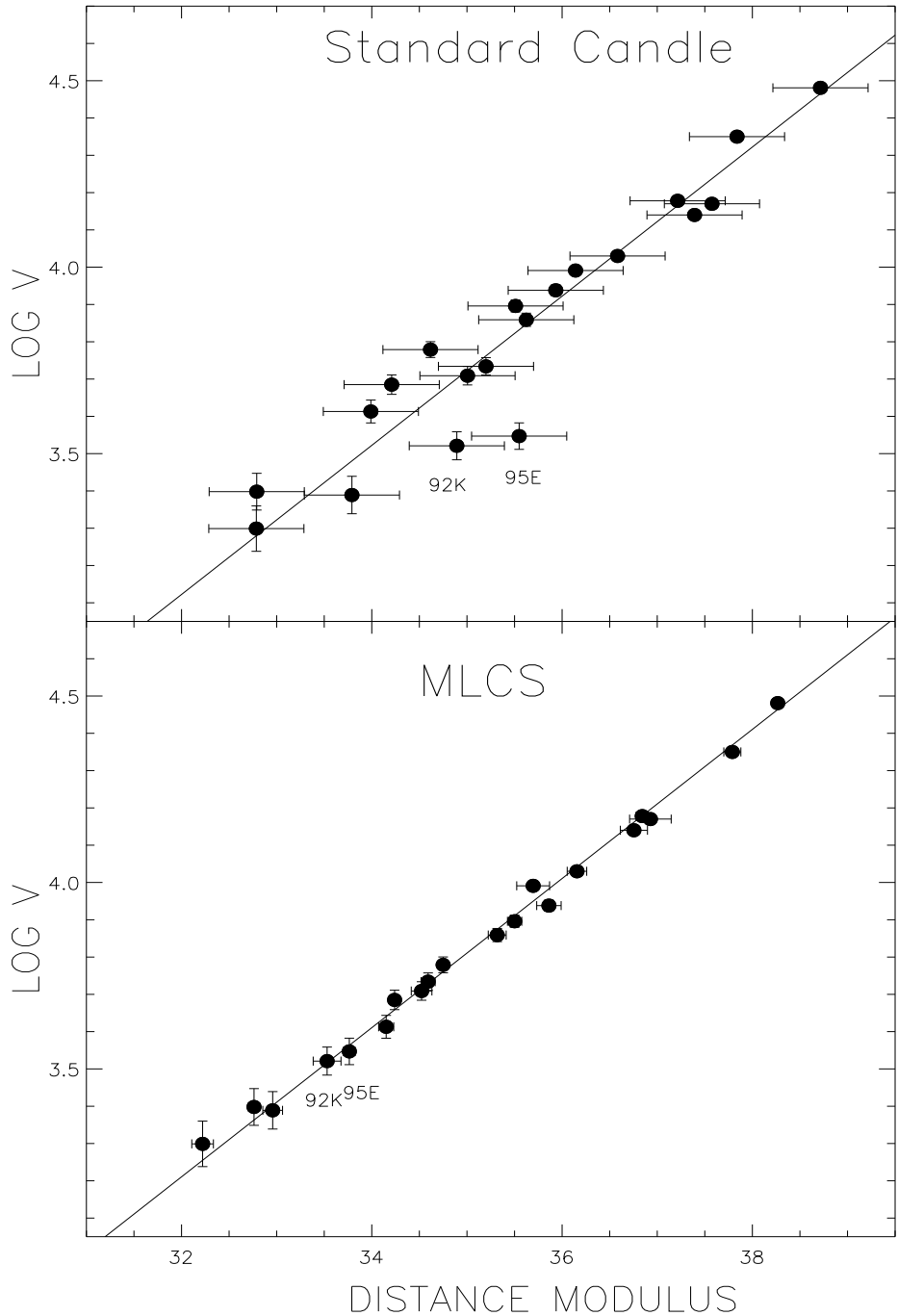}
\caption{Hubble Diagrams for SN Ia with velocities in the COBE rest frame
 on the Cepheid distance scale (Sandage et al 1994, 1996). All velocity errors are 300  km  s$^{-1}$ reflecting a plausible estimate of random velocities with respect to the Hubble
flow. (a) Distances estimated with a standard luminosity assumption and no correction for extinction.  This method yields $\sigma_v$=0.52 and H$_0$=$52 \pm 8$ (statistical) km  s$
^{-1}$  Mpc$^{-1}$(b) Distances from the MLCS method which makes a correction for intrinsic luminosity variation and total extinction as determined from the light and color curve shapes. This method yields $\sigma_v$=0.12 and H$_0$=$65 \pm 3$ (statistical) km  s$^{-1}$  Mpc$^{-1}$.}
\end{figure}
\eject

    \begin{table}                              
\begin{center}             
\caption{SN Ia distance comparisons (CMB frame, $\sigma_{vel}=300$ km s$^{-1}$)} \vspace{0.4cm}
\begin{tabular}{lllll} 
 \hline \hline
 $ \ \ \ \ \ $Correction& Form & all SN Ia, N=20 & w/o SN1995E, N=19  & color cut, N=18   \\
  luminosity $ \ \ \ \ A_V$  &    &   $\sigma \ \ \ \ \ \ \ P_r$ & $\sigma \ \ \ \ \ \ \ P_r$ & $\sigma \ \ \ \ \ \ \ P_r$ \\
\hline 
 none $ \ \ \ \ \ \ \ $ none & $\mu+\Delta$ & 0.52 & 0.40  & 0.33 \\
 MLCS  $ \ \ \ \ \ $ none & $\mu$ & $0.46 \ \ 0.015$ & $0.20 \ \ < 1$x$10^{-6} $ &$ 0.20 \ \ 2$x$10^{-5}$    \\
 MLCS $ \ \ \ \ \ $ galactic & $\mu-A_{V,gal} $ & $0.46 \ \ 0.25$ & $0.17 \ \ 0.018$  & $0.16 \ \ 0.0085$ \\
 MLCS $ \ \ \ \ \ $ MLCS & $\mu-A_V$ & $0.12 \ \ 1$x$10^{-5} $ & $0.13 \ \ 7$x$10^{-5}$  & $0.13 \ \ 0.00013$ \\
\hline
\hline
\end{tabular}
\end{center} 
\end{table}

    Moving down table 4, each successive row represents a refinement in our distance measuring technique.  The result of successive improvements in the method can be seen in the decreasing dispersion on the Hubble diagram.  Including the heavily reddened SN 1995E in the sample demonstrates the power of MLCS in dealing correctly with reddened objects but masks the gradual improvement in distance to be made from various aspects of the method.    It is important to note that even the distances of the color cut sample can be improved with MLCS.  This shows that the distance precision of ``normal'' or ``Branch-normal''  SN Ia can be significantly enhanced using light and color curve information.  The MLCS method is not just ``reining in'' the extreme SN Ia, but rather is improving the distance measures to all of the SN Ia.  

    To demonstrate the statistical significance of each level of improvement, we have included in table 4 the probabilities, $P_r$,  that the observed improvement in dispersion could occur from a {\it random} set of distance corrections.  These probabilities give the likelihood for the null hypothesis; that our distance ``corrections'' have no relation to the true SN Ia distances.  In a Monte Carlo simulation we apply a set of random corrections chosen from the distribution of proposed corrections and see how often the dispersion is as low or lower than the actual improved dispersion.  The value of $P_r$ gives the probability that the observed (or a greater) decrease in dispersion for each successive distance refinement occured by chance.  The results show MLCS luminosity and extinction corrections are highly significant regardless of the sample criterion.   Both the MLCS luminosity correction and extinction correction strongly reject the null hypothesis that they are unrelated to the true distance of the SN Ia.  Even the small corrections for galactic extinction do more good than harm.  Yet, accounting only for the Milky Way's contribution to the total absorption fails to account for host galaxy extinction which in a few cases can be substantial.  Using all the predictive power of MLCS gives remarkably low values for the dispersion with an exceedingly small probability that this improvement in dispersion occured by chance.  With a conservative estimate for the peculiar velocity associated with each field galaxy of 300 km s$^{-1}$ our observed dispersion of 0.12 magnitudes implies a typical distance precision of 5\%.  The improvement in distance precision by including a correction for host galaxy extinction with the conventional reddening law is the first demonstration that such corrections can be successfully made.

   Our Monte Carlo simulation demonstrates that even for a set of SN Ia selected by color, such as those used by Sandage et al (1996), the MLCS method makes a significant improvement in the precision of the distances.  For the color cut sample, the probability that both our luminosity and extinction corrections would improve the dispersion from 0.33 mag to 0.13 mag by chance is {\it less than one in a million}.

 Both SN 1992K and SN 1995E provide instructive examples of how this method leads to improved distance estimates.  Both objects appear to be dim, which for a standard candle, suggests they are at a great distance.  Assuming a standard candle luminosity places each of them much further away than their redshift implies (see figure 6a).  Yet, there are clues in the light and color curves of these objects which indicate that they are dim for different reasons.   The rapidly declining V light curve of SN 1992K and the color evolution of its B--V curve are nearly identical to the photometric behavior of the subluminous SN1991bg, a member of the training set (Hamuy et al 1994).  Application of the MLCS method estimates SN 1992K to be $\Delta=$1.25 magnitudes dimmer than the standard SN Ia, though its $A_V$ is only 0.01 mag. This correction to the luminosity is independent of the SN's redshift since it depends only on the light curve shape, but it is reassuring to note that accounting for this object's intrinsic faintness shifts its distance quite precisely onto the Hubble line.
  
   For the case of SN 1995E, the dim appearance that places this SN Ia below and to the right of the Hubble line  is not intrinsic to the supernova, but rather, is a result of absorption, as shown by MLCS.  This supernova was found on the spiral arm of NGC 2441.  The shape of its light and color curves suggests a fair resemblance to the standard SN Ia event ($\Delta=-0.17$ mag), but all its measured colors are systematically displaced to the red, as would occur from absorption by dust.   By fitting the {\it shape} of the color curves independently from the {\it value} of the color, we can measure the color excesses.  Assuming a standard reddening law (see Riess, Press, and Kirshner 1996b) we estimate the visual band extinction to be 1.86 magnitudes.  As in the case of SN 1992K, correcting the luminosity of SN 1995E makes its distance consistent with its redshift measurement.  Using a color cut requiring  $-0.25 \leq (B-V)_{max} \leq 0.25$, both of these objects would be discarded.  Yet, we can keep these objects (and others like them) in the sample and use the MLCS method to distinguish between supernovae which are intrinsically dim and those which are dimmed by dust absorption.

      We can make two significant checks of our extinction-corrected distances by examining the fit of the Hubble line to the data.    We examine the linearity of the Hubble law by measuring the slope of the relation between the MLCS distance modulus and the logarithm of the redshift.  Assuming that space is Euclidean for our modest redshifts, the expectation of this slope is 0.2.  Lauer \& Postman (1992) found a slope of $0.1992 \pm 0.006$ using brightest cluster galaxies and Jerjen \& Tammann (1993) found a slope of $0.1988 \pm 0.006$ using the mean of a number of distance indicators to fifteen clusters. Using the extinction-corrected distances and errors of table 3 yields a slope of $0.2010 \pm 0.0035$ which is consistent with 0.2 and with the two previous results.  The small error on our slope over the distance interval 32.2 $< \mu <$ 38.3 makes this the most precise check of this classical test of cosmology.  Finally, we can examine the goodness of fit of our Hubble line to the extinction-corrected MLCS distances.  The $\chi^2$ of the fit using the independently determined distance errors in table 3 and $\sigma_{vel}$ = 300 km s$^{-1}$ is 13 for 19 degrees of freedom, which is within the expectation of $\chi^2$.   The value of $\chi^2$ is strongly dependent on the assumed random velocity of the field galaxies hosting our SN Ia due to our low dispersion and small distance errors.   A random velocity error for our field galaxies in the range $125$ km s$^{-1} \leq \sigma_{vel} \leq 300$ km s$^{-1}$ gives a $\chi^2$ within its likely range of 13 to 25 and is consistent with other determinations of this random velocity component (Marzke 1995, Davis \& Peebles 1983).   All indications suggest that the MLCS method provides remarkably precise, extinction-corrected distances which are a significant improvement over SN Ia distances from previous methods.
  
\section{Discussion}    
Our intent has been to describe the MLCS method in enough detail so others can take advantage of the precise distance estimates it provides.  The strength of MLCS lies in its ability to disentangle the effects of absorption and intrinsic luminosity variation while providing meaningful error estimates.   These measures are derived from the distance independent observables of multicolor light curve shapes.  The accuracy of the MLCS {\it relative} distance measures has been well established on an independent set of twenty SN Ia on the Hubble diagram. 
    
 To place our MLCS distances on the established {\it absolute} distance scale we use the luminosity calibration for a number of SN Ia with an independent distance indicator of high precision.   At present, there are three SN Ia whose light curves meet our MLCS quality criteria (modern photoelectric photometry with observations less than ten days after maximum) and whose distances have been measured with Cepheids observed with the Hubble Space Telescope.   
 \begin{table}                              
\begin{center}             
\caption{SN Ia distance calibration} \vspace{0.4cm}
\begin{tabular}{lccccc} \hline \hline 
SN Ia & $\mu_{MLCS}(\sigma)$ & $\mu_{Ceph}(\sigma)$ & $\mu_{MLCS}-\mu_{Ceph}(\sigma)$ & $\Delta(\sigma)$ & $M_V-M_{V,81B}(\sigma)$ \\
\hline
1972E  & 27.21$^*$(0.09) &  28.08$^*$(0.10)  & 0.87(0.13) &  -0.33(0.04) & -0.28(0.21) \\
1981B  & 30.38(0.07) & 31.10(0.20) & 0.72(0.21) & 0.01(0.04) & 0.00(0.00) \\
1990N & 31.04(0.14) & 32.00(0.23) & 1.01(0.0.27) & -0.27(0.04) & -0.07(0.23) \\
\hline
\hline
\multicolumn{5}{l}{ \scriptsize \ \ $^*$ distance modulus uncorrected for extinction} \\
\end{tabular}
\end{center}
\end{table}

The three SN Ia, SN 1972E, SN 1981B, and SN 1990N, are listed in table 5: column (2) gives the extinction-corrected MLCS distances on the distance scale of SBF-PNLF-TF (table 1), column (3) gives best distances as determined by HST Cepheid measurements (Sandage et al 1994, 1996), column (4) gives the differences between columns (3) and (2), column (5) gives the MLCS luminosity correction, and column (6) gives the Sandage et al (1996) $M_V$ minus the $M_V$ for the ``standard'' shaped SN 1981B.  By comparing our precise MLCS {\it relative} distances to the trustworthy Cepheid {\it absolute} distances we see a consistent difference of $0.81 \pm 0.10$ mag.  Adding this difference to all of our distance estimates places our distance indicator on the Cepheid distance scale.  Further, it provides an absolute luminosity calibration for the standard ($\Delta=0$) SN Ia of $M_V=-19.36 \pm 0.10$ mag at B maximum.  This calibration for the standard $M_V$ template is used in table 2. The extinction-corrected distances listed in table 3 include this offset which places them on the Cepheid distance scale.  We can calculate the Hubble constant by fitting \bq log v = .2(\mu-A_V) + log \ H_0 - 5 \eq to the distances and velocities in table 3 using the tabulated errors $\sigma_{\mu-A_V}$ and an assumed random velocity of 300 km s$^{-1}$.  The result is a Hubble constant of 66 $\pm$ 3 km s$^{-1}$ Mpc$^{-1}$ where the uncertainty is internal and incorporates a 0.03 mag uncertainty in the Hubble line and a 0.10 mag uncertainty in the placement onto the Cepheid distance scale.  We also see that the MLCS luminosity corrections (column 5) are consistent with the luminosity differences as determined by the Cepheids (column 6).

 \begin{table}                                 
\begin{center}             
\caption{SN Ia Hubble constant comparisons (CMB frame, $\sigma_{vel}=300$ km s$^{-1}$)} \vspace{0.4cm}
\begin{tabular}{llccc} \hline \hline 
 $ \ \ \ \ \ $Correction& Form & all SN Ia, N=20 & w/o SN1995E, N=19  & color cut, N=18   \\
  luminosity $ \ \ \ \ A_V$  &    &   $H_0$ & $H_0$ & $H_0$ \\
\hline
  none $ \ \ \ \ \ \ \ $ none  & $\mu+\Delta$ & $52 \pm 8$ & $54 \pm 6$  & $55 \pm 5$ \\
 MLCS  $ \ \ \ \ \ $ none& $\mu$ & $61 \pm 9$ & $63 \pm 4$ &$62 \pm 4$    \\
 MLCS $ \ \ \ \ \ $ galactic  & $\mu-A_{V,gal} $ & $63 \pm 9$ & $65 \pm 4$  & $64 \pm 4$ \\
  MLCS $ \ \ \ \ \ $ MLCS  & $\mu-A_V$ & $65 \pm 3$ & $65 \pm 3$  & $65 \pm 3$ \\
\hline
\hline
\end{tabular}
\end{center}
\end{table}

  For comparison we perform the same calculation for the other distance and extinction methods using the three SN Ia samples.  We apply each method to the calibration set to determine the offset onto the Cepheid scale, and to the distant set to determine the Hubble line.  For the methods which do not provide their own error estimates, we assume a constant distance error of the size necessary to get the expected $chi^2$ of the fit to the Hubble line.  In table 6 we list the determinations of the Hubble constant and internal error for the different methods and samples.

The color cut sample provides us with a set of SN Ia which are directly comparable to the ``normal'' set used by Sandage et al (1996).  Using the assumption of a standard luminosity and the color cut sample gives $H_0=55 \pm 5$ km s$^{-1}$ Mpc$^{-1}$ which is consistent with the Sandage et al (1996) values of $H_0(B)=56 \pm 4$ and $H_0(V)=58 \pm 4$ km s$^{-1}$ Mpc$^{-1}$  obtained with the same standard brightness assumption and sample criteria, though with a different set of light curves.  

    In \S 6 and table 4 we demonstrated that each of the distance measuring improvements en route to the complete MLCS method is highly significant and should be employed to get the best result.   Using the MLCS method to measure extinction-corrected distances and errors yields a Hubble constant of $H_0=65 \pm 3$ km s$^{-1}$ Mpc$^{-1}$ or $H_0=63 \pm 3$ km s$^{-1}$ Mpc$^{-1}$ using constant weighting.  The only significant change in the Hubble constant arises from including the luminosity correction, $\Delta$.  This is because the mean luminosity of the three nearby calibrators is $\sim$ 15\% dimmer than the mean luminosity of the twenty SN Ia in the distant sample.  That these nearby SN Ia are dimmer than the distant SN Ia {\it does not} imply an anti-selection bias because our sample is neither volume nor magnitude limited.  If our sample was complete in volume or magnitude, we would observe a tremendous increase in the number of observed SN Ia at large distances.  Figure 6 shows this is clearly not the case.  We elaborate on this point below.

A complete error budget for any of the values of $H_0$ in table 6 would consist of the stated internal error added in quadrature to the estimated uncertainty in the Cepheid zero point, roughly 0.15 mag (Feast and Walker).  For the MLCS extinction-corrected distances, this gives a value of $H_0=65 \pm 6$ km s$^{-1}$ Mpc$^{-1}$.  

     The MLCS distances to SN Ia in the Hubble flow combined with redshifts provide the necessary information to estimate the peculiar velocity component of the each objects' radial motion.  Plotted on the sky, the velocity residuals show a dipole pattern indicative of the motion of the Local Group with respect to a frame defined by the supernovae.  Our preliminary analysis of this motion with a subset of 13 SN from the current set and without our MLCS reddening information was consistent with convergence to the cosmic microwave background frame and inconsistent with the Lauer \& Postman frame (1994) at 7000 km s$^{-1}$ (Riess, Press, Kirshner 1995b).  The same analysis performed at higher precision using the independent set of twenty SN Ia and the MLCS method (which now includes extinction corrections) yields an even stronger detection of the Local Group motion with similar results.  In the future, when the sample of SN Ia has grown, we will revisit our analysis of the Local Group motion from SN Ia.

   Sections 2-5 outline the MLCS technique in sufficient detail to allow for future redetermination of the necessary templates and functions as more supernovae become available or are deemed desirable to include in the training set.  The training set for this paper uses all supernovae for which accurate and extensive photoelectric photometry is currently available as well as definitive relative distance estimates and $A_V$ estimates.  These included from the Phillips (1993) set, SN 1980N, SN 1981B, 1986G, 1989B, 1990N, 1991T, 1991bg, and 1992A.  We exclude SN 1971I for which only photographic photometry is available .  We also use SN 1994ae (Riess et al 1996) for which there are detailed photometric light curves and an independent distance (Dell'Antonio 1995).   Ideally we would use only SN Ia's with no evidence of extinction, but currently we lack a sufficient number of training set objects to discard any.  Instead, we estimate $A_V$ from the color differences of unreddened SN Ia's of similar light curve shape and propagate the resulting uncertainty in $A_V$ to our distance errors. 

     Richmond et al (1995) have discussed a conspicious inconsistency with a preliminary SBF distance to NGC 4526 (John Tonry, private communication) and the distance from the shape of the SN 1994D light curve.  The difference of 0.7 magnitudes for the SBF and supernovae distance moduli to NGC 4526 imply a ``greater than 3-sigma'' mutual rejection.  The case can be simplified by comparing the results for SN 1992A in NGC 1380 in Fornax to SN 1994D.  These supernovae have virtually identically shaped light curves, but SN 1994D peaked 0.65 magnitudes brighter than SN 1992A implying that it was 33\% closer.  SBF suggests these two supernovae are in galaxies at similar distances (Phillips 1993).  Current HST Key project observations of Cepheids in Fornax may eventually help to ``break the tie''.  

    Inclusion of SN 1994D has the effect of decreasing the precision of the distance estimates for the independent sample of objects, but the training set is large enough that the effect is not substantial.   The dispersion of distances on the Hubble diagram for MLCS increased from 0.12 mag to 0.22 mag and the $\chi^2$ increased from 13 to 39.  If SN 1994D were truly at the SBF distance and was representative of objects in the independent set we would expect to see no significant change in the Hubble diagram $\chi^2$ for including this object.  If however, the SBF distance is incorrect or this one object is a special case, then we would expect to see a substantial change in the value of $\chi^2$ as we have.  A final possibility would be that SN 1994D is representitive of a departure from the behavior of other SN Ia and it has only been due to chance that we have not seen any others in the independent sample.  In this case, the increased dispersion would be a more accurate measure of the true dispersion of well-behaved SN Ia's and ``94D-like'' objects treated as a whole.  Future discovery of such ``SN 1994D-like'' objects could lead to a new class of objects at least for the purposes of distance measurement.  Until either another such object is observed or the SBF distance is finalized or an independent distance is measured we remain agnostic.    

    In section 5 we developed a formal way to combine our measurement of $A_V$ with our {\it a priori} understanding of dust.  This method requires some description of the distribution of $A_V$ values for supernovae discovered in galaxies.  The most conservative estimate for $p(A_V)$ is that it is constant over the range of equation (22).   This amounts to truncation of extinctions less than zero.  A gaussian wing which is a maximum at $A_V$=0.0 and is parameterized by its second moment seems more plausible.  Either assuming a constant value for $p(A_V)$ or a gaussian wing with 0.5 $\leq$ $\sigma$ $\leq$ $\infty$ yields a high level of distance precision.  The Hubble diagram dispersion is insensitive to the value of $\sigma$ over this range, and the Hubble constant is insensitive to any particular parameterization of $p(A_V)$ including requiring $A_V$=0 (see table 6).  Using a $p(A_V)$ with $\sigma$ $\leq$ 0.5 is too restrictive and amounts to discarding any extinction information.  Determination of the best value for $\sigma$ by maximizing the log-likelihood of equation (15) on the Hubble diagram yields $\sigma$=0.9 mag.  It is important to consider sample selection effects when choosing a form for the observed $p(A_V)$.  Clearly the width of $p(A_V)$ could be a function of the search characteristics since these determine how likely it is to find an SN Ia obscured by dust.  Monte Carlo calculations provide an estimate of the effect of misjudging $p(A_V)$, and with enough data, one could easily solve for the distribution of observed $A_V$.  

Tammann and Sandage (1995) have questioned the validity of our training set distances; ``Supernovae used by Phillips (1993) ...based on distances determined by the Tully-Fisher and the surface brightness fluctuation methods...clearly deviate from the [Cepheid calibrated] supernovae values."   This criticism is misdirected.  While it is true that the {\it absolute} distance scale as determined by Cepheid variables is not necessarily consistent with the Tully-Fisher and SBF distances, we only employ {\it relative} distances of these methods to measure the {\it relative} luminosity variation of SN Ia out to the Virgo cluster (Kennicutt, Freedman, \& Mould 1995).   The selection bias pointed out by Sandage (1988a,b 1994a,b) to distort Tully-Fisher distances {\it does not apply} to the distances in our training set which were derived from the bias-corrected inverse Tully-Fisher relation (Strauss \& Willick 1995, Schechter 1980, Pierce 1996).  

 Tammann and Sandage (1995) have also commented on the statistical problem raised by a light curve-luminosity relation stating that ``the Malmquist effect on the distant SN Ia would be overwhelming''.  Specifically, Tammann and Sandage (1995) argue that the distant SN Ia occupy ``a volume of about 3 x $10^5$ larger, on average, than the volume of the three nearby local calibrators''.  Therefore, one would expect a substantial selection bias favoring brighter SN Ia in the distant sample as compared to the SN Ia nearby.  In fact, the mean SN Ia luminosity for the distant sample we use is $0.34$ magnitudes {\it dimmer} than for the set of three nearby calibrators.   Is this surprising?  Not very.  The selection bias, as stated by Tammann and Sandage (1995), is estimated for a complete sample of SN Ia with a well-defined limiting magnitude.  This is clearly not case for our sample or for theirs.  Inspection of the Hubble diagrams of figure 6 (or of Tammann \& Sandage's) does not show the number of SN Ia increasing with distance as $10^{0.6\mu}$ as expected for a complete search of increasing volume.  This point can be made quantitatively by performing a simple $\langle {V \over V_{max}} \rangle$ test (Schmidt 1968); which has an expectation value of 0.5 for a uniformly distributed sample.  For the distant set of 20 SN Ia $\langle {V \over V_{max}} \rangle$=0.09, which shows that this sample is concentrated nearby.  Even limiting the test to the set of ten SN Ia discovered during the uniform Calan/Tololo survey yields $\langle {V \over V_{max}} \rangle$=0.16.  Real samples of observed SN Ia are not distributed the way assumed by Tamman \& Sandage (1995).  The chances of the three local calibrators being $0.34$ magnitudes dimmer than the independent sample is 6 \%.  This is about as likely as randomly picking the winner of the next baseball American League Championship series (ALCS).
 
An interesting question to consider is what explosion or progenitor parameters could explain our empirically determined variation in light curve shape, luminosity and color?  Current
 models have attempted to explain the inhomogenity of supernovae in one of two ways. 
H\"{o}flich, Khokhlov, \& Wheeler (1995) and H\"{o}flich \& Khokhlov (1996)
have found that a variation in the density at which the deflagration burning front
 transitions to a detonation wave affects the amount of $^{56}$Ni produced in the explosion.   A
   late transition gives the outer layers time to preexpand resulting in a small nickel production. The reduced nickel  heating diminishes  the temperatures in the expanding envelope and photosphere. This results in  rapidly dropping  opacities. Consequently,
  the photosphere recedes fast and the stored energy is emitted over a short period of time. Conversely, an early turnover of the deflagration into a detonation front results in a large amount
  of nickel which produces a bright and hot supernova whose opaque layers keep the radiative energy loss comparably low. This mechanism works for deflagrations, delayed detonations and pulsating delayed detonations. Qualitatively,  the observed correlation between luminosity and the photometric parameters are reproduced (H\"oflich \& Khokhlov 1996).   

     Another theoretical approach to matching the observations involves exploding progenitors of varying and generally sub-Chandrasekhar mass.   A layer of helium accumulates at the base of the static hydrogen-burning zone.  Sudden burning of this helium at the base of the layer sends a shock that can trigger a carbon detonation at the interface.  If this does not occur, a second chance at carbon detonation can come when the shock propagates around the star and converges on the opposite side.  The variation in progenitor mass suggests a simpler connection to the amount of nickel produced than the Chandrasekhar models.  Again the variation in nickel yield is expected to match the observed correlations in supernova observables.  These models have been successful in one and two dimensions, but it remains to be seen if the abundance and the velocity distribution of the intermediate mass elements produced from nucleosynthesis matches the observations (Livne and Arnett 1995, Woosley \& Weaver 1994, Livne 1990, Livne \& Glasner 1990)

   Finally, improvements which can be made in the MLCS technique when we acquire a larger training set of supernovae.  First, the ``grey snakes'' of figure 3 and the relative weights of the B,V,R, and I data determined from  equation (16) suggest that greater precision could be attained by including a quadratic term in the light curve shapes model for B,R, and I band data.  An additional order in the fit would demand a larger training set of data to avoid overfitting the details of the training set objects.  A larger training set would also support the determination of a more detailed correlation matrix with estimates of model residual covariances.

     The MLCS technique provides an exceedingly precise way of measuring extinction-corrected distances with uncertainty estimates.  The redshifts to which Type Ia supernova light curves can measure distances have interesting implications for cosmological measurements.  Two teams are currently laboring to find and measure SN Ia at 0.3 $\leq$ z $\leq$ 0.6 (Perlmutter et al 1996, Schmidt et al 1996) with the intent of measuring the cosmological deceleration of the Universe, $q_0$.  At these redshifts SN Ia light curves are difficult to obtain, and informative spectra even harder, so it is sensible to use all the available SN Ia.  The uncertainty in $q_0$ is proportional to the distance precision divided by the square root of the number of objects.  The MLCS method should help with both of these factors by including all well-observed SN Ia in the sample, and increasing the precision of SN Ia distance measures.

We are again grateful to Mario Hamuy, Mark Phillips, Nick Suntzeff and the entire Cal\'{a}n/Tololo collaboration for the opportunity to study their exceptional data before publication.  We have greatly benefited from discussions with Brian Schmidt, George Rybicki, and Peter H\"{o}flich.  This work was supported through NSF grants AST 92-18475 and PHY 95-07695.
\eject

\def\refitem{\par\parskip 0pt\noindent\hangindent 20pt}
\centerline {\bf References}
\vskip 12 pt

\refitem Arnett, W.D.,Branch, D.,Wheeler, J.C.,1985, Nature 314,337
\refitem Arnett, W.D., 1969, {\it Astrophys. Space Sci.}, 5, 280
\refitem Baade, W. 1938, ApJ, 88,285
\refitem Baade, W. \& Zwicky, F., 1934, Proc. Nat. Acad. Sci., 20, 254
\refitem Barbon, R.,Ciatti,F., and Rosino, L., 1973, A\&A 25,65

\refitem Branch, D. and Tammann, G.A.,1992, ARA\&A, 30,359
\refitem Branch, D., 1992, ApJ, 392, 35
\refitem Branch, D. and Miller, D. 1993, ApJ, 405, L5
\refitem Branch, D., Fisher, A., \& Nugent, P., 1993, AJ, 106, 2383
\refitem Boisseau, J.R., Wheeler, J.C., 1991, AJ, 101, 1281
\refitem Burstein, D., \& Heiles, C. 1982, AJ, 87, 1165

\refitem Buta, R.J., \& Turner, A., 1983, PASP, 95, 72
\refitem Capaccioli, M., Cappellaro, E., Della Valle, M., D'Onofrio, M., Rosino, L., \& Turatto, M. 1990, ApJ, 350, 110
\refitem Chapman, Giles, 1988, AJ, 95, 999
\refitem Ciardullo, R., Jacoby, G.H., Tonry, J.L., 1993, ApJ, 419, 479
\refitem Colgate, S. \& McKee,  1969, ApJ, 157, 623
\refitem Cristiani, S. 1992, A\&A, 259,63
\refitem Curtis, H.D. 1921, BNRCNAS, 2, 171
\refitem  de Vaucouleurs, G. et al. 1991, in {\it  Third Reference Catalogue of Bright Galaxies} (Springer-Verlag, New York)
\refitem Davis, M., \& Peebles, P.J.E., 1983, ApJ, 267, 465
\refitem Della Valle, M., \& Panagia, N. 1992, AJ, 104,696
\refitem Dell'Antonio, I., 1995, private communication
\refitem Doggett, J.B., Branch, D. 1985, AJ, 90, 2303
\refitem Filippenko, A.V. et al. 1996, in preparation
\refitem Filippenko, A.V. et al. 1992, AJ, 104, 1543
\refitem Ford, C. et al. 1993, AJ, 106, 3
\refitem Goldhaber, G., et al., 1996, in preparation
\refitem Hamuy, M., Phillips, M.M., Maza, J., Suntzeff, N.B., Schommer, R.A., \& Aviles, A.  1996,in preparation
\refitem Hamuy, M., Phillips, M.M., Maza, J., Suntzeff, N.B., Schommer, R.A., \& Aviles, A. 1995, AJ, 109, 1 
\refitem Hamuy, M., Phillips, M.M., Maza, J., Suntzeff, N.B., Schommer, R.A., \& Aviles, A. 1994, AJ, 108, 2226
\refitem Hamuy, M., et al. 1993a, AJ, 106, 2392
\refitem Hamuy, M., Phillips, M., Wells, L., \& Maza, J. 1993b, PASP, 105, 787
\refitem Hamuy, M., et al., 1991, AJ, 102, 208
\refitem Hoyle, F., \& Fowler, W.A., 1960, ApJ, 132, 565
\refitem H\"{o}flich, P., Khovkhlov, A., \& Wheeler, J.C. 1995, ApJ, 444, 831
\refitem H\"{o}flich, P., \& Khovkhlov, A., 1996, ApJ, 457, 500
\refitem Humason, M.L.,, Mayall, N.U., \& Sandage, A.R., 1956, AJ, 61, 97
\refitem Jacoby, G.H., et al. 1992, PASP, 104, 678
\refitem Joeever, M., 1982, {\it Astrofizica}, 18, 574
\refitem Kennicutt, R.C., Freedman, W.L., \& Mould, J.R., AJ, 110, 1476
\refitem Khovkhlov, A.,Muller,E. and H\"{o}flich, P. 1993, A\&A, 270,223

\refitem Kirshner, R.P., Oke, J.B., Penston, M.V., \& Searle, L. 1973, ApJ, 185, 303
\refitem Kowal, C. T. 1968, AJ, 73, 1021
\refitem  Lauer, T.R. \& Postman, M. 1994,ApJ, 425,418

\refitem  Lauer, T.R. \& Postman, M. 1992, ApJ, 400, L47
\refitem Leibundgut, B. et al. 1996, submitted to ApJ
\refitem Leibundgut, B. et al. 1993, AJ, 105,301
\refitem Leibundgut, B. et al. 1991, ApJ, 371, L23
\refitem Leibundgut, B. 1989, PhD Thesis, Basel
\refitem Lira, P., 1995, Masters Thesis, University of Chile
\refitem Livne, E., Arnett, D., 1995, ApJ, 452, 62
\refitem Livne, E. 1990, ApJ, 354, L53
\refitem Livne, E., \& Glasner, A.S., 1990, ApJ, 361, L244
\refitem  Lynden-Bell, D. \& Lahav, O. 1988, {\it Large-Scale Motions in the Uni
verse} (ed. Rubin, V.C. \& Coyle, G.V.) (Princeton, Princeton University Press)
\refitem  Marzke, R.O., Geller, M.J., daCosta, L.N. \& Huchra, J.P. 1995, AJ, 110, 477
\refitem Maza, J., Hamuy, M., Phillips, M., Suntzeff, N., Aviles, R. 1994, ApJ, 424, L107
\refitem Mazurek, T.J., Wheeler, J.C., 1980, {\it Fund. Cosmic Phys.}, 5, 193
\refitem Miller, D. and Branch, D. 1990, AJ, 100,530

\refitem Minkowski, R. 1941, PASP, 53, 224
\refitem Minkowski, R. 1964, ARA\&A, 2, 247
\refitem Nomoto, K., Thielemann, F., Yokoi, K. 1984, ApJ, 286,644
\refitem Oke, J.B., \& Sandage, A., 1968, ApJ, 154, 21
\refitem Oke, J.B., \& Searle, L., 1974, AR\&A, 12, 315
\refitem Perlmutter, S. et al., 1996, NASI, Aiquablava, Spain
\refitem Perlmutter, S. et al., 1995, ApJ, 440, 41
\refitem Phillips, M. 1993, ApJ, L105
\refitem Phillips, M., Wells, L., Suntzeff, N., Hamuy, M., Leibundgut, B., Kirshner, R.P., Foltz, C.,  1992, AJ, 103, 1632

\refitem Phillips, M. et al., 1987, PASP, 99,592
\refitem Pierce, M.J., 1996, BAAS, 1293
\refitem Pierce, M.J., 1994, ApJ, 430, 53
\refitem Press, W.H., Teukolsky, S.A., Vetterling, W.T., and Flannery, B.P. 1992, Numerical Recipes, 2ed (Cambridge University Press)

\refitem Pskovskii, Y. 1984, {\it Sov. Astron.}, 28,658
\refitem Pskovskii, Y. 1977, {\it Sov. Astron.}, 21,675

\refitem Richmond, M.W., Treffers, R.R., Filippenko, A.V., Van Dyk, S.D., Paik, Y., \& Peng, Chien, 1995, AJ, 109, 2121   
\refitem Riess, A.G., et al., 1996, in preparation
\refitem Riess, A.G., Press W.H., Kirshner, R.P., 1996b, in preparation
\refitem Riess, A.G., Press W.H., Kirshner, R.P., 1995a, ApJ, 438m L17
\refitem Riess, A.G., Press W.H., Kirshner, R.P., 1995b, ApJ, 445, L91
\refitem Rood, H.J., 1994, PASP, 106, 170
\refitem Ruiz-Lapuente, P. et al., 1993, Nature, 365, 728
\refitem Rybicki, G. \& Press, W., 1992,ApJ, 398,169
\refitem Rybicki, G. \& Kleyna, J.T., 1994, in {\it Reverberation Mapping of the Broad-Line Region in Active Galactic Nuclei} ed P.M. Gondhalekar, K. Horne, \& B.M. Peterson, PASP
\refitem Sandage, A. et al. 1996, ApJ, 000, 000
\refitem Sandage, A. et al. 1994, ApJ, 423, L13
\refitem Sandage, A., \& Tammann, G.A., 1993, ApJ, 415, 1
\refitem Sandage, A., Tammann, G.A., Panagia, N., \& Macchetto, D., 1992, ApJ, 401, 7L
\refitem Sandage, A., 1994a, ApJ, 430, 1
\refitem Sandage, A., 1994b, ApJ, 430, 13
\refitem Sandage, A., 1988a, ApJ, 331, 583
\refitem Sandage, A., 1988b, ApJ, 331, 605
\refitem Savage, B.D. \& Mathis J.S., 1979, ARA\&A, 17, 73
\refitem Schaefer, B.E., 1996, submitted for publication
\refitem Schaefer, B.E., 1995a, ApJ, 447, L13
\refitem Schaefer, B.E., 1995b, ApJ, 449, L9
\refitem Schaefer, B.E., 1994, ApJ, 426, 493
\refitem Schecter, P.L., 1980, AJ, 85, 801
\refitem Schmidt, B.P., et al., 1996, in preparation
\refitem Schmidt, B.P.,  Kirshner, R.P. \& Eastman, R., 1992, ApJ, 395, 366
\refitem Schmidt, M., 1968, ApJ, 151, 393
\refitem Shapley, H. 1921, BNRCNAS, 2, 171
\refitem  Smoot, G.F. et al.. 1992, 396, L1

\refitem Strauss, M.A., \& Willick, J.A., 1995, PhR, 1995, 261, 271
\refitem Suntzeff, N.B., et al., in preparation
\refitem Suntzeff, N.B., 1993, in {\it IAU Colloquium 145, Supernovae and Supernovae Remnants}, ed R. McCray, Cambridge University Press
\refitem Tammann, G.A., \& Sandage, A., 1995, ApJ, 452, 16
\refitem Tammann, G.A., \& Leibundgut, B. 1990, A\&A, 236, 9
\refitem Tammann, G.A., 1987, in {\it IAU Symposium 124, Observational Cosmology}, ed A. Hewitt, G. Burbidge, L.Z. Fang, p. 151, Dordrecht: Reidel
\refitem Tonry, J.L., 1991, ApJ, 37
\refitem Tonry, J.L., private communication
\refitem Uomoto, A. \& Kirshner, R.P., 1985, A\&A, 149, L7
\refitem van den Bergh, S., 1995, ApJ, 453, L55
\refitem Vaughan, T.E., Branch, D., Miller, D.L., \& Perlmutter, S., 1995, ApJ, 439, 558
\refitem Wells, L.A., et al., 1994, AJ, 108, 2233
\refitem Wheeler, J.C., \& Levreault, R., 1985, ApJ, 294, L17  
\refitem Wheeler, J.C., Harkness, R.P., Barkat, Z., Swartz, D., 1986, PASP, 98, 1018
\refitem Wheeler, J.C., \& Harkness, R.P., 1990, Rep. Prog. Phys.,53,1467
\refitem Woosley, S.E. and Weaver, T.A. 1994, ApJ, 423, 371

\refitem Woosley, S.E., and Weaver, T.A., 1992 in: Supernovae
\end{document}